\documentclass[12pt,preprint]{aastex}

\shorttitle{Superflare on II~Peg}
\shortauthors{Osten et al.}

\begin{document}

\title{ Nonthermal Hard X-ray Emission and Iron K$\alpha$ Emission
from a Superflare on II~Pegasi
}
\author{Rachel A. Osten\altaffilmark{1}}
\affil{Astronomy Department, University of Maryland, College Park, MD 20742
rosten@astro.umd.edu}
\altaffiltext{1}{Hubble Fellow}

\author{Stephen Drake\altaffilmark{2}, Jack Tueller, Jay Cummings}
\affil{NASA Goddard Space Flight Center, Greenbelt, MD 20771}
\altaffiltext{2}{Also USRA}

\author{Matteo Perri}
\affil{ASI Science Data Center, Via Galileo Galilei, I-00044 Frascati, Italy}

\author{Alberto Moretti and Stefano Covino}
\affil{INAF-Osservatorio Astronomico di Brera, via Bianchi 46, I-23807 Merate, Italy}

\begin{abstract}
We report on an X-ray flare detected on the active binary system
II~Pegasi with the Swift telescope.  The event triggered the Burst Alert Telescope
in the hard X-ray band on December 16, 2005  at 11:21:52 UT with
a 10-200 keV luminosity of 2.2$\times$10$^{32}$ erg s$^{-1}$ 
--- a superflare, by comparison with
energies of typical stellar flares on active binary systems.
The trigger spectrum indicates a hot thermal plasma with T$\sim$180 $\times$10$^{6}$K.
X-ray spectral analysis from 0.8--200 keV with the X-Ray Telescope and BAT
in the next two orbits reveals
evidence for a thermal component (T$>$80 $\times$10$^{6}$K) and Fe K 6.4 keV emission.
A tail of emission out to 200 keV can be fit with either an
extremely high temperature thermal plasma (T$\sim$3$\times$10$^{8}$K)
or power-law emission.  Based on analogies with solar flares, 
we attribute the excess continuum emission to nonthermal thick-target
bremsstrahlung emission from a population of accelerated electrons.
We estimate the radiated energy from 0.01--200 keV to be $\sim$6$\times$10$^{36}$ erg,
the total radiated energy over all wavelengths $\sim$10$^{38}$ erg, the energy
in nonthermal electrons above 20 keV $\sim$3$\times$10$^{40}$ erg, and conducted
energy $<$5$\times$10$^{43}$ erg.
The nonthermal interpretation gives a reasonable value
for the total energy in electrons $>$ 20 keV when compared to the upper and lower
bounds on the thermal
energy content of the flare.
This marks the first occasion in which evidence exists for
nonthermal hard X-ray emission
from a stellar flare.  
We investigate the emission  mechanism responsible for producing the 6.4 keV feature,
and find that collisional ionization from nonthermal electrons appears to be more plausible
than the photoionization mechanism usually invoked on the Sun and pre-main sequence stars.
\end{abstract}


\section{Introduction}
The general model for solar (and stellar) flares involves a release of free energy
via magnetic reconnection, which accelerates particles and causes subsequent plasma motions
and heating \citep{dennis1989}.  
Particle acceleration during solar flares can be diagnosed directly through
nonthermal hard X-ray bremsstrahlung emission and radio gyrosynchrotron emission.
Until now, evidence for particle acceleration during stellar flares has been
either indirect \citep[through proxies such as optical and UV flare emissions;][]{slh1995,slh2003}
 or achieved through
the detection of nonthermal radio flare emissions \citep{klein1987}, 
under the assumption that the observed
radiation is gyrosynchrotron emission from accelerated particles. 
The interpretation of radio observations, 
while useful for constraining the characteristic magnetic field in the source, is complicated by
optical depth effects and spatial inhomogeneities in the emitting source, which are usually
not uniquely determined.
Previous detections of hard X-ray emission from stellar flares 
could not demonstrate unequivocally the presence of 
nonthermal emission; instead, a superhot thermal component could equally 
well explain the observed 10--50 keV emission 
\citep{fs1999,pallavic2001,franciosini2001} without the need for any additional power-law
components.
The detection and characterization of nonthermal hard X-ray emission during stellar flares
would be important, as it would allow a direct investigation of the energy spectrum of accelerated electrons,
and a comparison of the total thermal/nonthermal energy budget.  
Prospects for detecting such emissions have normally been taken to be remote, as during solar flares the 
nonthermal hard X-ray emission (20 keV-- 1 MeV) is $>$10$^{-5}$ times less intense than soft X-ray (1--10 keV)
emission \citep[see, e.g., the composite solar flare spectrum in Fig.~94 of ][]{aschwanden2002}; 
this is a consequence
of collisional energy losses exceeding bremsstrahlung losses by a factor of 10$^{5}$.
This necessitates large flare sizes for hard X-ray detections, which are intrinsically 
rare, which coupled with a dearth of pointed
observations exacerbates the situation.

Iron K fluorescence emission has been seen during solar flares \citep{parmar1984,zarro1992}
and also seen in impulsive flares from some pre-main sequence stars \citep{coupref}.  The origin
on the Sun
is attributed to photoexcitation of K shell electrons in photospheric iron 
by X-ray bremsstrahlung radiation.  The emission mechanism for pre-main sequence stars
is the same as for the Sun, but the fluorescence is presumed to arise instead from X-ray irradiation
of an accretion disk rather than the stellar photosphere.  There have not been any previous reports
of iron K$\alpha$ fluorescent emission from non-accreting stellar flare sources.
Such an identification would provide another
link between solar flare physics and the physics of flares on other stars.

II~Peg (HD~224085) is a well-known active binary system which displays evidence for starspots covering a large
fraction \citep[$>$40\%;][]{marino1999} of the primary's photosphere, and has signatures 
associated with vigorous magnetic activity across the entire electromagnetic spectrum.
At a distance of 42 pc \citep{hipparcos}, this single-lined spectroscopic binary
 is composed of a 0.8 M$_{\odot}$ K2IV primary 
tidally locked in a 6.7 day orbit with a $\sim$0.4M$_{\odot}$ secondary \citep{berd1998}.
Space velocities indicate that this binary is a member of 
the old disk stellar population \citep{eggen1978}.  Both photospheric and coronal
metal abundances have been determined, leading to an apparent depletion of coronal iron by
about a factor of 4 relative to the photospheric value \citep[][and references therein]{chandraref}
\footnote{This is in contrast with the solar corona, where the coronal iron abundance appears
to be {\it enhanced} by a factor of $\sim$4 over the solar photospheric abundance
\citep{feldmanlaming2000}.  For more discussion on stellar coronal abundances, see \citet{screview}.}.  
Based on the stellar parameters for II~Peg in \citet{berd1998} and using the bolometric
corrections in \citet{bcref}, we estimate the bolometric luminosity, the power emitted by the
two stars over all wavelengths, 
to be $L_{\rm bol}\approx$5.5$\times$10$^{33}$ erg s$^{-1}$.

II Peg was first detected as a flaring X-ray source by the {\it Ariel-V}
Sky Survey Instrument (1.5-20 keV) in the mid-1970s
\citep{schwartz1981}
which detected 2 flares with peak fluxes of $1.1 \times 10^{-9}$ and
$1.5 \times 10^{-9}$ erg cm$^{-2}$ s$^{-1}$ in the 2-10 keV band,
equivalent to peak X-ray luminosities of $1.6 \times 10^{32}$
and $2.8 \times 10^{32}$ erg s$^{-1}$. 
Since then, moderate and large flares from II Peg have been detected
by many subsequent X-ray and EUV observatories, including {\it Ginga} \citep{doyle1991},
{\it EXOSAT} \citep{tagliaferri1991}, {\it EUVE} \citep{patterer1993,ostenbrown},
{\it ASCA} \citep{mewe1997},
{\it BeppoSAX} \citep{covino2000}, and {\it Chandra} \citep{chandraref}.

In this paper, we report on a large flare observed on II~Peg by the detectors on the
Swift Gamma-Ray Burst Mission \citep{Swiftref}.
Following the discussion in \citet{schaefer2000} and the characteristics of the flare
(radiated energy, luminosity) we term this event a ``superflare''.
The event triggered Swift's Burst Alert Telescope (BAT)
in the hard X-ray band on December 16, 2005 at 11:21:52 
UT\footnote{http://gcn.gsfc.nasa.gov/gcn/gcn3/4357.gcn3}, and its characteristics
in the soft and hard X-ray bands were determined for the extent of time for which hard
X-ray emission was detected, $\approx$ 7000 s.  
The UV$/$Optical Telescope did turn on during the trigger, but the images were saturated due to II~Peg's 
brightness (V$\approx$7.4), and the data were of little use.  In this paper we concentrate on analysis of the
X-ray Telescope (XRT) and BAT spectra.

\section{Data Reduction}
\subsection{BAT Data Reduction}
BAT is a coded-mask instrument with a very large ($\sim$1.2 sr) field of view.
The construction and operation of the instrument is discussed in
\citet{BATref}. An overview may be found at:
http://swift.gsfc.nasa.gov/docs/swift/about\_swift/bat\_desc.html.
All Swift data are made public as soon as possible, and the raw BAT
data are available by navigating from the above site.


The BAT data used in this analysis are ``survey" data, intended for
the BAT all-sky hard X-ray survey. It consists of accumulated counts,
in time bins from 60 to 300 seconds, for each of the ~32k CZT
detector elements, in 80 energy bins nominally from 10 to 194 keV plus high
and low integral energy bins. The survey data are supplemented during
an interval around the BAT trigger time, including during the Swift
slew maneuver, by ``event" data, which contain higher resolution
time and energy data on each photon count in the detector array.

We used BAT pipeline software within FTOOLS\footnote{The FTOOLS software package
provides mission-specific data analysis procedures; a 
full description of the procedures mentioned here can be found at http://heasarc.gsfc.nasa.gov/docs/software/lheasoft/; the names of individual procedures referred to in this and the following section are 
italicized.} version 6.0.3 to correct the energy from the efficient
but slightly non-linear energy assignment made on board. For the
spectral data reported here, we used {\it batbinevt} to produce mask-weighted
spectra in several broad time intervals. For the
light curve data we created sky images in two broad energy bins for each
time interval using {\it batbinevt} and {\it batfftimage}, and found the flux at the
source position using {\it batcelldetect}, after removing a fit to the diffuse
background and the contribution of bright sources in the field of view.

\begin{figure}[h]
\begin{center}
\includegraphics[scale=0.5,angle=90]{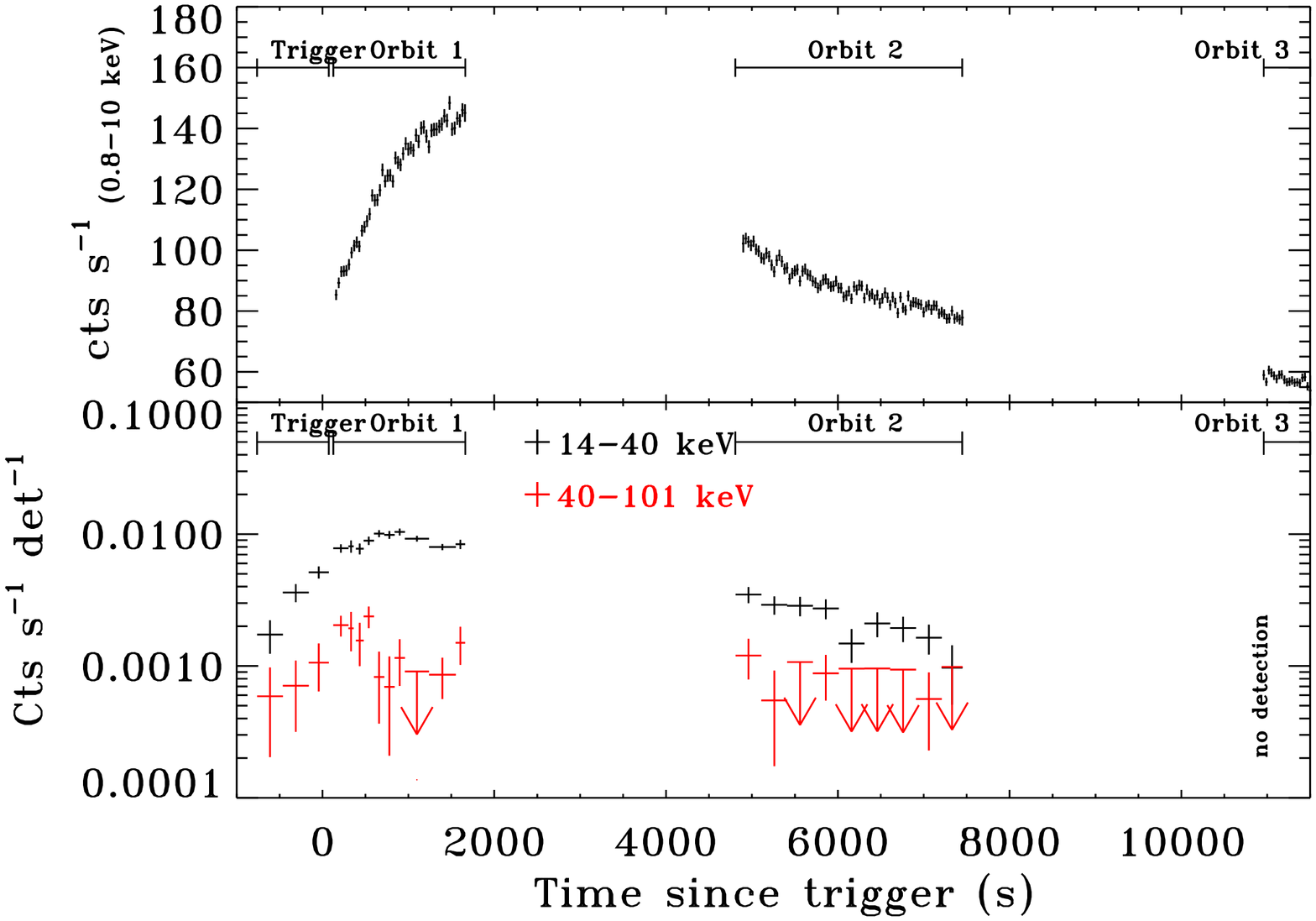}
\figcaption[]{Light curves from XRT, 0.8--10 keV (top) and two hard X-ray energy bands from the
BAT (14--40, 40--101 keV;
bottom).  Time bins where SNR$<$1 are shown as downward arrows. 
Time is expressed referenced to the trigger of 11:21:52 UT on 16 December 2005.\label{fig:lc}}
\end{center}
\end{figure}

As seen in Figure~\ref{fig:lc} , the spectral data are reported in 3 intervals relative
to the BAT trigger time of 11:21:52 UT: ``Trigger", T-760 to T+126 sec;
``Orbit 1", T+126 to T+1661 sec; and ``Orbit 2", T+4810 to T+7451 sec. There
was no significant flux in the BAT energy range during Orbit 3.  To increase the signal-to-noise ratio
at energies above 38 keV, the BAT spectra were grouped by 3 bins.

\subsection{XRT Data Reduction}
For a technical description of XRT and its operations, we refer the reader to \citet{Burrows_XRT}.
XRT started observing at 11:24:16 UT, that is, 144~s after the BAT trigger.
XRT observed IIPeg in three different snapshots in three following orbits.
The first snapshot (``Orbit 1") lasted 1518 s.  The second snapshot (``Orbit 2") started at 12:43:28 and
lasted 2545~s. The last snapshot (``Orbit 3") began at 14:24:16 UT and lasted 535~s.
To produce the cleaned and calibrated event files, the data were reduced by
means of the {\it xrtpipeline} task 
and calibration files of the CALDB~20051221 release\footnote{A description and history of
calibration files for Swift can be found at http://swift.gsfc.nasa.gov/docs/heasarc/caldb/swift/}.
Since the source count rate was over 60 counts per second during 
the entire observation all data were collected in windowed timing (WT) mode.
In order to extract the spectrum and light curve, events were selected
with grade 0--2\footnote{See discussion in \S 2.3 of http://swift.gsfc.nasa.gov/docs/swift/analysis/xrt\_swguide\_v1\_2.pdf for classification of events and grades.} from a 20 pixel (47\arcsec) box corresponding to 80\%
of the encircled energy fraction (EEF) at 1.5~keV.
We restricted our analysis to the energy band 0.8-10 keV, ignoring
lower energy channels.
In fact, the effective area calibration files (ancillary response functions, or ARFs)
included in the CALDB~20051221
release were still preliminary and present systematics in the energy range
0.2-0.8 keV, which are noticeable in this high intensity event.
Background extraction regions were chosen to be the same shape as the source 
extraction region, at approximately 50 pixels
from the center of the source.
The mean count rate in the source region in the three orbits was 161, 118, and 79
counts per second respectively, whereas the background count rate never
exceeded 0.5 counts per second.
In order to study the spectral variations of the Fe K line (see below),
we split the first orbit in three segments (144--650, 650--1156,1156--1662 s from the trigger),
and the second in two (4902--6174,6174--7447~s from the trigger).
The light curve was produced using the standard task {\it lcurve} and with a temporal bin
of 30~sec. The spectra were grouped with the task {\it grppha} in order to have 20 counts
as a minimum for each energy bin. The ancillary response function was produced by means of the
{\it xrtmkarf} taking into account the point-spread function correction \citep{Moretti_SPIE}.

\section{Spectral Analysis}
Figure~\ref{fig:lc} shows the 0.8--10 keV XRT light curve and BAT light curves 
in two energy bands.  
It is evident that the harder X-ray energies have progressively
earlier peaks:  at 0.8--10 keV, the light curve peaks $\sim$1480 s following the BAT trigger,
while, for the 14--40 keV band, the peak is $\sim$840s post trigger, and for the 40-100 keV band,
the maximum
count rate occurs only $\sim$480 s after the trigger.  This points to the hardest X-ray photons having
a more impulsive behavior than the soft X-ray photons.
To examine the origin and evolution of these hard X-rays, we examined spectra corresponding to four different time intervals: the trigger spectrum and data collected in spacecraft orbits 1, 2 and 3.
Spectral fitting was performed
using XSPEC (v12.0)\footnote{XSPEC, an X-ray spectral fitting package,
is part of the Xanadu software package, which is released packaged with the FTOOLS described
previously and found at the same website.}.
To describe the line and continuum emission from a thermal, diffuse, collisionally ionized plasma,
we used a customized version of the APEC model \citep{apecref}\footnote{For more information
see http://cxc.harvard.edu/atomdb/} which has continuum
emission calculated out to
photon energies of 100 keV.
Spectral fitting proceeds via a forward method, computing model spectra with varying
input parameters, convolving these with the instrument resolution and sensitivity  and
comparing them with 
observed spectra until suitable statistical convergence is achieved.
We allow elemental metal abundances to vary in fitting each XRT spectrum, 
but because of the low spectral resolution in the soft X-ray region,
the abundances are scaled to a common multiple $A$
of the solar abundance (e.g. for a best-fit scaled abundance $A=$0.5, the
Fe/H is 2$\times$Fe/H$_{solar}$,
and the Mg/H is also 2$\times$Mg/H$_{solar}$).
We use the solar photospheric composition of \citet{gs1998} in our spectral fitting; all 
fitted abundances
quoted in this paper are with respect to these values.  In particular, the $Fe/H$
ratio in \citet{gs1998} is 3.16$\times$10$^{-5}$.

There are deviations of the model from the data at low energies; the most prominent
of these is a feature near 1.09 keV which appears in all XRT spectra.
The residual near 1.09 keV looks very similar, both in amplitude and shape, to those of the 
Crab spectrum used to calibrate the XRT.  This feature is most
likely an instrumental effect similar to other features observed in the (excluded)
energy range below 0.8 keV.
There are other large residuals around 1 keV; these may indicate that, 
instead of the Ne abundance scaling with
the Fe abundance (in which case, the Ne/Fe abundance ratio would be one with respect to the solar
ratio, which is 3.8), the Ne/Fe
abundance ratio is greater than this value.  Alternatively, it could also indicate the presence of
a lower temperature component, which we could not constrain, as its major contributor is at
energies $\leq$1 keV.

We have also included photoelectric absorption in the spectral fit, with the column density $N_{H}$
as a free parameter.  The best-fit values do not show significant variation,
but are systematically higher than previously determined interstellar column densities towards
II~Peg \citep[$\sim$5$\times$10$^{18}$ cm$^{-2}$;][]{mewe1997,patterer1993}:
this discrepancy is likely due to calibration uncertainties in the XRT at low energies.

\subsection{Trigger Spectrum}
\begin{figure}[h]
\begin{center}
\includegraphics[scale=0.3,angle=270]{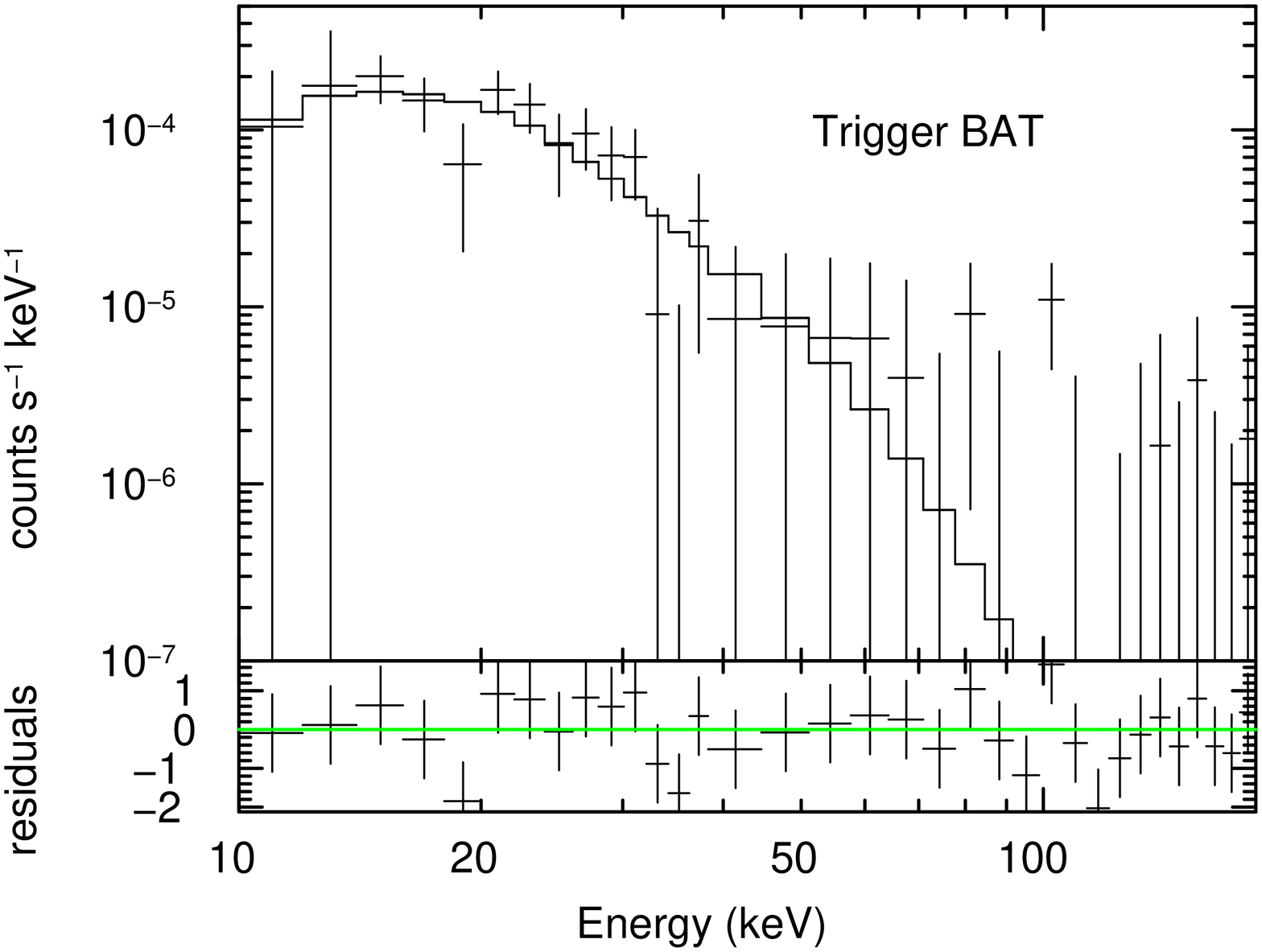}
\includegraphics[scale=0.3,angle=270]{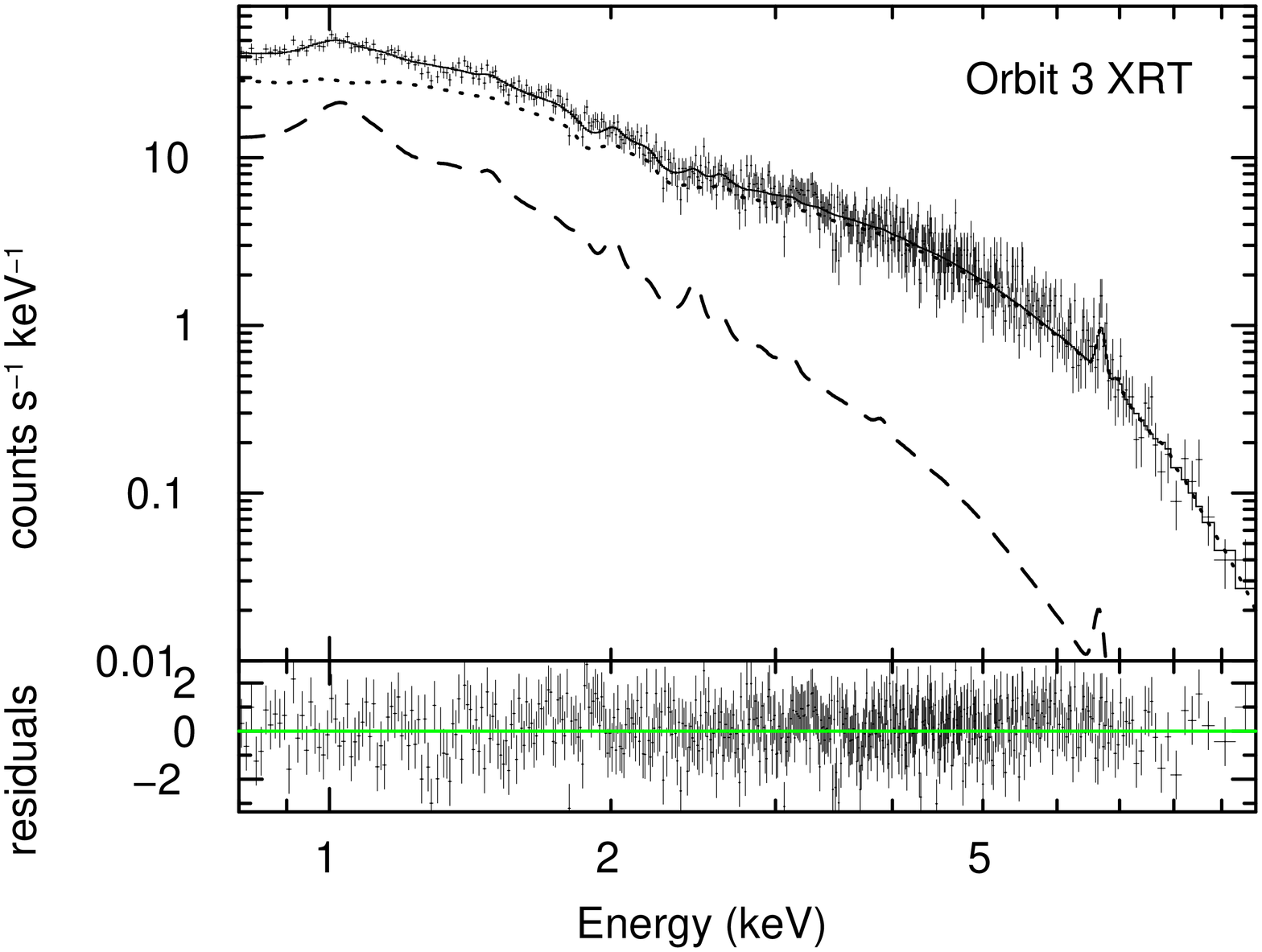}
\figcaption[]{Spectral fit to BAT trigger spectrum (left) and XRT spectrum in Orbit 3 (right).
For the trigger spectrum a single temperature component is fitted, while for Orbit 3
two temperature components are fitted.
Data with error bars are indicated by crosses. Histogram shows final model.
For Orbit 3 the contribution of the lower temperature component is shown with a dashed line,
while the contribution of the higher temperature component is shown with a dotted line.
Residuals are differences between data and model in units of $\sigma$.
Spectral fits are tabulated in Table~\ref{tbl:xspec}.
\label{fig:trigger}}
\end{center}
\end{figure}

The BAT trigger spectrum could be fit with a single component model, either a thermal (APEC)
model with abundances fixed to solar \citep{gs1998},
or a power-law model.  
At the high temperatures returned by the APEC model ($\sim$180 $\times$10$^{6}$K) continuum emission dominates.
Both models
fit the data statistically equally well; however,
we favor the thermal model based on previous behavior in large stellar flares 
\citep{fs1999,pallavic2001,franciosini2001} which behavior indicate the presence of high-temperature thermal
plasma.  Spectral fit results are given in Table~\ref{tbl:xspec} and shown in Figure~\ref{fig:trigger}.

\subsection{Orbits 1 and 2}
We investigated the presence of large residuals around 6.4 keV in the XRT spectrum during
Orbits 1 and 2, and added a line at this energy, fixing the energy at 6.4 keV and width at
the instrumental value.
The right bank of panels in Figures~\ref{fig:spec1} and ~\ref{fig:spec2} display 
the 5.5--9 keV region in Orbit 1 and Orbit 2 which includes this feature.
We interpret this excess to be the signature of K$\alpha$ emission from neutral or low ionization
states of iron, and we defer the discussion of its
origin and interpretation to \S4.5.

Only in two time intervals (Orbits 1 and 2) were both XRT and BAT spectra available, 
covering the 0.8--200 keV
energy range with significant signal to noise.  
Our spectral fitting for these times started with
a single thermal component, adding in additional thermal components to reduce the fit statistic.
The large value of the second temperature component can be confirmed by examination
of the 5.5--9 keV region (right bank of panels in Figures~\ref{fig:spec1} and ~\ref{fig:spec2})
which contains several line diagnostics sensitive to 40--160 $\times$10$^{6}$
K plasma.  The He-like and H-like Iron transitions at $\sim$6.7 and $\sim$6.9 keV are the
most prominent, and their ratio constrains there to be plasma at $\approx$ 10$^{8}$K.
Since the single temperature fit was not adequate to the spectra in either Orbit 1 or 2,
we do not discuss it further.
Temperatures near 10$^{8}$ K have also been inferred from previous large stellar flares 
\citep[e.g.,][]{abdor,franciosini2001}.
Figure~\ref{fig:spec1} shows the spectral fitting results in Orbit 1,
and Figure~\ref{fig:spec2} shows the spectral fitting results for Orbit 2.
The left bank of panels displays the fits to the 0.8--200 keV energy range for
two temperature (2T+G), three temperature (3T+G), and two temperature plus nonthermal (2T+G+NT)
models
(all have a Gaussian component at 6.4 keV as described above). 
The residuals plotted are deviations of the model from the data, calculated in units of $\sigma$.
The two-temperature fit has several energy bins above 40 keV (Orbit 1)
and 20 keV (Orbit 2) with significant residuals, which prompted us to add in the additional
component.  

In the case of three thermal components, the highest temperature component is modelled using
a thermal bremsstrahlung model whose contribution in the spectrum appears mainly at high energies.
The normalization of the bremsstrahlung component is given in $n_{e}n_{I}dV$, as compared
to the APEC model normalization $n_{e}n_{H}dV$, $n_{I}$ being the ion number density and $n_{H}$ the hydrogen
number density.
For both orbits, the $\chi^{2}_{\nu}$ statistic is minimized with three thermal components, but the
temperature of the highest component, $\approx$ 300 $\times$10$^{6}$K, in both Orbit 1 and 2, is very high,
and we reject this model for the
following reasons.
At high temperatures, the conductive cooling time dominates.  We can compute the ratio of
the thermal relaxation time of the plasma to the timescale for conductive cooling:\\
\begin{equation}
\tau_{\rm relax}/\tau_{\rm cond} = 200 \frac{T_{8}^{4}}{n_{10}^{2}L_{9}^{2}}
\end{equation}
\citep[see discussion in ][]{benzbook}, where $T_{8}$ is the temperature in units of 10$^{8}$ K,
$n_{10}$ is the electron density in units of 10$^{10}$ cm$^{-3}$, and $L_{9}$ is the 
loop length in units of 10$^{9}$ cm.   
For T$_{8}$=3, the timescale over which
the plasma can relax to a thermal distribution at this temperature exceeds by a large factor
the timescale on which the plasma would lose its energy via conductive losses, unless
the density is very high and/or the length scales involved are very long.
At high densities, the magnetic field required to confine the plasma becomes
large, 
B$_{conf}$=$\sqrt{8\pi n_{e}k_{B}T}=60\sqrt{n_{10}T_{8}} G$.
At low densities, the length scales exceed the primary stellar radius ($\sim$3R$_{\odot}$)
and the binary separation \citep[$a\sin i=$3.4$\times$10$^{11}$cm,][]{cabs}.
There have been numerous discussions in the literature concerning the thermal or nonthermal nature
of hard X-ray emission from solar flares \citep{vilmer1987,brownsmith1980}.
Because other lines of evidence exist which confirm that particle acceleration does occur
during stellar flares (mostly microwave gyrosynchrotron emission), we consider a
model which describes bremsstrahlung emission from suprathermal electrons.
The arguments which apply to 3$\times$10$^{8}$ thermal plasma apply also to the second
temperature component found in Orbits 1 and 2 (T$_{2}$ from 1.2--1.5 $\times$10$^{8}$
in Orbit 1, T$_{2}$ from 7--9 $\times$10$^{7}$ K in Orbit 2) with
only a slight easement of the physical restrictions described above.  
However, the line information present in 
the He-like and H-like iron transitions confirm the existence of this plasma, whereas the 
3$\times$10$^{8}$ component comes from continuum emission at hard X-ray energies.

Thus, we consider an additional model which contains a
nonthermal component.  This is based on hard X-ray emission from solar flares, which
usually shows evidence of bremsstrahlung radiation from a population of
nonthermal electrons \citep{dennis1989}.
If the emission arises as the result of a beam of 
injected electrons propagating downward through the atmosphere, we expect the observed radiation
spectrum to differ from the injection spectrum, being modified by collisions with
the increasingly dense atmosphere.  The formulation for describing such ``thick-target''
hard X-ray emission is described by \citet{brown1971}, where the observed spectrum $I(\epsilon)$
has the form \\
\begin{equation}
I(\epsilon) \sim \epsilon^{-\gamma} \;\;\; photons \;\;\; cm^{-2}\;\;\; s^{-1} \;\;\; keV^{-1}
\end{equation}
where $\epsilon$ is the photon energy and $\gamma$ is the photon index.
Under the formalism of thick-target bremsstrahlung emission, the observed spectrum can
be related to the spectrum of injected electrons, $F(E)\sim E^{-\delta}$ erg s$^{-1}$
keV$^{-1}$, 
where $E$
is the electron energy, and $\delta$ is the power-law index of the electron distribution
function; $\delta$ is related to $\gamma$ as $\delta=\gamma+1$.
We implemented 
a thick-target bremsstrahlung code\footnote{Based on the BREMTHICK code developed for
analysis of RHESSI solar flare data available at 
http://hesperia.gsfc.nasa.gov/hessi/modelware.htm} in XSPEC to model the 
behavior of the hard X-ray photons.  
The spectral shape is determined by the power-law index of the electron injection spectrum; the
normalization depends on the total power input in accelerated electrons and
the low-energy cutoff in the electron injection spectrum.
We cannot constrain the cutoff with our dataset, and fix this value to 
an arbitrary value of 20 keV often used in solar flare analysis \citep{dennis1989}.
Recent RHESSI solar flare results \citep{holman2003} have 
determined 37 keV as the highest value consistent with their data,
although much lower energy cutoffs (down to a few keV) have also been determined \citep{sui2006}.
This value (20 keV) is roughly comparable with the approximate location in the spectrum where the nonthermal
spectrum starts to dominate.
Note that the choice of low energy cutoff has dramatic consequences for the power input in 
accelerated electrons: decreasing the cutoff to a lower value $E_{c}$ increases the 
power input necessary to reproduce the spectrum by a factor $(20/E_{c})^{\delta-2}$, or
a factor $\approx$ two higher for $\delta \sim 3$ and $E_{c} =10$ keV.


The fact that the 3rd temperature component in Orbit 2 is also unphysically high, coupled with
the large residuals above 20 keV between model and data for a two-temperature fit and the
$\chi^{2}_{\nu}$ statistic for the 2 temperature plus nonthermal model being nearly
the same as for the three temperature model,
leads us to conclude that there is evidence for nonthermal emission into the decay
phase of the flare.
Indeed, since the thermal plasma has cooled ($T_{2}\approx$ 8$\times$10$^{7}$K in Orbit 2
compared with T$_{2}\approx$1.4$\times$10$^{8}$K in Orbit 1), the discrepancy between model and
data for the two-temperature model shows up at smaller X-ray energies (see Figure~\ref{fig:spec2}).
Based on either the three temperature model or the two temperature plus nonthermal model,
in Orbit 1 the 0.8--10 keV flux is 6.12$\times$10$^{-9}$ erg cm$^{-2}$ s$^{-1}$ [(6.09--6.13)$\times$10$^{-9}$ 1$\sigma$ uncertainty], 
and the 10--200 keV flux is 3.9$\times$10$^{-9}$ erg cm$^{-2}$ s$^{-1}$ [(3.6--4.0)$\times$10$^{-9}$ 1$\sigma$ uncertainty].  
Expressed as a fraction of the bolometric luminosity (L$_{\rm bol} \approx$5.5$\times$10$^{33}$ erg s$^{-1}$), 
these values are 0.24 (0.8--10 keV) and
0.15 (10--200 keV).
In Orbit 2,
the 0.8--10 keV flux is 3.62$\times$10$^{-9}$ erg cm$^{-2}$ s$^{-1}$ [(3.61--3.64)$\times$10$^{-9}$
1$\sigma$ uncertainty], and 10-200 keV flux is 1.29$\times$10$^{-9}$ erg cm$^{-2}$ s$^{-1}$ [(1.11--1.37)$\times$10$^{-9}$ 1$\sigma$ uncertainty].
Expressed as a fraction of the bolometric luminosity, these values are
0.14 (0.8--10 keV) and 0.05 (10--200 keV).

\subsection{Orbit 3}
In the third orbit following the trigger, there was no BAT detection of the source.  The XRT
did collect data for an interval of $\sim$ 535 s, which we analyzed.  There is no statistical evidence indicating
the need for either a power-law component at the highest energies, or for a 6.4 keV feature.  The
thermal component is still at a relatively high level, and the 0.8--10 keV flux is $\sim$40
times higher than recorded during quiescent intervals seen with Chandra \citep[5.2$\times$10$^{-11}$
erg cm$^{-2}$ s$^{-1}$, 0.5--12 keV;][]{chandraref}. 
We conclude that the flare in soft X-rays lasted longer than the $\sim$12 ks in
which it was observed by Swift. Spectral fit results are listed in Table~\ref{tbl:xspec}
and the data and model are plotted in Figure~\ref{fig:trigger}.

\section{Discussion}
The soft X-ray flux peaked during Orbit 1, with the radiated energy flux $\gtrsim$ 100 times
that observed during previous observations made when the system was quiescent
\citep{chandraref}.  Thus, we apply the
``superflare'' moniker to describe this event.
There have been previous detections of stellar flares in the hard X-ray band of 20--50 keV
\citep{fs1999,pallavic2001,franciosini2001,integral}, yet
none have shown unambiguous evidence for nonthermal emission.  Indeed, 
detailed spectral fitting for these flares has indicated the presence of a superhot thermal
plasma component (T$>$80 $\times$10$^{6}$K), but, with no detections above $\sim$50 keV there 
has been no evidence
of a power-law tail in the spectrum due to nonthermal emission.  These results were somewhat puzzling,
in that radio observations necessitate continuous acceleration of electrons to describe
the steady levels of microwave emission \citep{cdf1993}.  In addition, intense
radio flares reaching 2--3 orders of magnitude enhancement over the quiescent level of
microwave emission have been observed \citep{richards2003} which probably correspond
to the ``superflare'' classification
as well.

We estimate the peak flux during Orbit 1 as if this were a solar flare -- 
the 1--8 \AA (1.55--12.40 keV) flux at Earth would be 440 W m$^{-2}$ for $d=1AU$ instead
of $d=42 pc$. Using the notation for
X-class solar X-ray flares ($Xn=n\times$10$^{-4}$ Wm$^{-2}$), 
this would correspond to X4.4$\times$10$^{6}$.  
For comparison, the largest solar flare yet observed has been $\sim$X30.  Given the huge difference
in radiative output, we should not expect this flare to look and behave like solar flares.
The key features of this superflare are (1) its intensity in the soft X-ray band, notably
the evidence for high temperature thermal emission; (2)
evidence for nonthermal emission at hard X-ray energies; and (3) the presence of 6.4 keV
Iron K emission.  We discuss the implications of each of these for the physics of
flares in the following sections.

\subsection{High Temperature Thermal Emission and the Neupert Effect }
High temperature plasma ($>$ 5$\times$10$^{7}$K) is evident in all four time intervals.
The hottest component, $\approx$ 1.8$\times$10$^{8}$K, appears during the trigger, although
this result is based only on 10--200 keV continuum emission.
The spectral fitting to Orbits 1 and 2 involving models with two or more temperature components
returns a high temperature ($\sim$70--150 $\times$10$^{6}$K) which is generally hotter in Orbit 1
than in Orbit 2.
We have argued in a previous section (\S 3.2) that the temperature returned by modelling
the spectra in Orbits 1 and 2 with three temperature components is not as plausible as the
existence of a nonthermal thick-target bremsstrahlung emission component at hard X-ray energies, and
so we will ignore the three temperature fits in favor of the nonthermal interpretation.
The two temperature fit to the spectrum in Orbit 3 also reveals evidence for plasma at $\approx$
5$\times$10$^{6}$K.
Thus, the general trend appears to be that the plasma is cooling during this progression from the 
trigger through Orbit 3, as deduced by
the decrease in the hottest temperature component from these spectral fits, with an 
exponential decay time of $\sim$3 hours.

The correspondence in many solar flares between thermal coronal energies and the amount and
timescale of nonthermal energy deposition \citep[the ``Neupert Effect'';][]{neupert1968} lends credence
to the posited physical association between particle acceleration and coronal heating.
Stellar flares also sometimes show evidence of the Neupert effect, although typically radio emission
or other proxies for hard X-ray emission (e.g. optical white-light flares) are used to compare
with soft X-ray thermal radiation \citep{gudel1996,gudel2002,slh1995}.  
We have investigated the temporal relationship between
the soft X-ray emission and hard X-ray emission in this flare to see if it shows the Neupert effect.
In this case, the time rate of change of the thermal energy content of the corona
(as revealed by soft X-ray radiation)
should be roughly equivalent to the instantaneous nonthermal energy luminosity,
$\dot{E}_{th} \propto L_{NT}$.   
The 40--101 keV energy band of the BAT spectrum appears to be dominated by power-law
hard X-ray emission, so we use this light curve to describe the temporal variation in
nonthermal energy deposition.  The thermal coronal energy is released mainly in the 0.8--10 keV
energy band, so we use this light curve to constrain the temporal evolution in the thermal
energy.
Figure~\ref{fig:neupert} displays the correspondence between the derivative of the XRT light curve
and the instantaneous nonthermal hard X-ray emission.  The greatest correlation occurs during the first
part of Orbit 1. We take this as suggestive of a Neupert effect relationship between
the nonthermal hard X-ray emission and thermal coronal radiation, although there is complex
behavior here, as the nonthermal hard X-ray emission appears to persist into the decay phase of the flare.
An additional complication lies in whether the plasma heating to such high temperatures takes place
concurrently with, or as a result of, the particle acceleration producing the nonthermal emission.

The Neupert effect argues for a causal relationship between particle acceleration and plasma heating.
The energy dissipation which sets off the chain of events seen in a flare can heat plasma as well as
accelerate particles \citep{dennis1989}; the exact proportion depends 
on the partition between plasma heating and
particle acceleration, which is unknown.
It is possible that plasma heating to the high temperatures seen during the trigger and subsequent
orbits occurs before or concurrently with particle acceleration,
although we cannot constrain the presence of the nonthermal component during the time of the trigger, due
to poor signal-to-noise constraints, and we also
cannot deduce what the pre-trigger conditions were. 
\citet{benkaholman} discussed a model in which both thermal and nonthermal emissions occur
and are physically linked due to electron currents both heating plasma by Joule dissipation
and accelerating electrons via a runaway process.
Secondary heating such as predicted by the Neupert effect is then a different effect.
\citet{solarneupert} show that high temperature solar flare plasma ($T>16$$\times$10$^{6}$K) is more likely to
exhibit the Neupert effect than low temperature plasma, but this assumes that the
flare energy release occurs predominantly in nonthermal electrons.
\citet{li1993} computed hard and soft X-ray time profiles using models in which the
hard X-ray emission was produced by either a super-hot thermal component or 
nonthermal thick-target bremsstrahlung, and concluded that the thermal model
fails to reproduce the derivativity relationship between hard X-ray and soft X-ray emission.
A situation where primary plasma heating and secondary plasma heating due to energy lost by
nonthermal electrons would be more complicated.  
The current data appear to be consistent with a Neupert effect relationship, but a more
confident interpretation relies on knowing the times when nonthermal emisison and high
temperature flare emission appeared.

As discussed by \citet{feldman1996}, there appears to be a relationship between
flare temperature and emission measure for solar flares and large stellar flares; 
``bigger'' flares (more intense, larger emission
measure) tend to be hotter.  
The high temperature component of
the flare under consideration here also appears to fit with this general relation.
\citet{bigsolarflares} found a correlation between nonthermal hard X-ray flux and thermal 
plasma parameters
from a sample of solar flares, indicating that flares with large values of nonthermal emission
also have higher temperatures and emission measures.
Applying this result to the current flare confirms that 
a strong bias exists in detecting nonthermal hard X-ray emission
from stellar flares, as very large and energetic (and hence rare, due to the flare frequency-energy relationship) flares are needed to achieve detections of nonthermal emission at hard X-ray energies.  
This problem is
further compounded by the addition of the superhot thermal component, itself a consequence of
the large flare, which complicates detection of the nonthermal component in the hard X-ray
spectrum.

\subsection{Nonthermal Hard X-ray Emission}
The power $F_{0}$ in the accelerated electrons as deduced by
thick-target bremsstrahlung spectral fits is $\sim$ 10$^{37}$ erg s$^{-1}$, as listed in Table~\ref{tbl:xspec}.
As noted above, this number depends on the value of the low energy cutoff in the injected electron
spectrum, which we cannot constrain.  
However, our somewhat arbitrary choice of 20 keV as a lower energy cutoff does fit with the 
region of the spectrum where nonthermal emission begins to dominate.
The spectral indices of the electron distribution 
returned from the spectral fitting are $\delta$ of 2.8 and 3.1; these compare favorably with 
spectral indices inferred from solar HXR burst spectral indices \citep[e.g.,][]{mp1991}; 
significant evolution of the hard X-ray spectrum is usually observed during solar flares, but
our sensitivity constraints do not permit examinations on a finer time scale.

The flare lasted for $>$ 11000 s in the soft X-ray band, 
but the timescale for hard X-ray emission is shorter, $\sim$
7000 s, based on the nondetection by BAT in the third orbit following the trigger.  
The major energy loss mechanism for the nonthermal electrons will be via collisions with the
ambient thermal electrons; we thus
estimate the energy loss timescale as \\
\begin{equation}
\tau_{defl}= 9.5\times10^{7} s \left(\frac{E_{keV}^{3/2}}{n_{e}}\right) \left(\frac{20}{\ln \Lambda} \right) \;\;\; .
\end{equation}
where $E_{keV}$ is the energy of the electron in keV, $n_{e}$ is the ambient plasma density in the 
region of energy loss in cm$^{-3}$,
and $\ln \Lambda$ is the Coulomb logarithm, $\approx$10 under typical chromospheric conditions.
We use $n_{e}$ of $\sim$10$^{10}$--10$^{11}$ cm$^{-3}$ 
based on electron density measurements of II~Peg's lower
transition region/upper chromosphere given in \citet{doyle1992}. 
Under these conditions, a 100 keV electron will lose its energy in 2--20 s.  This is significantly
shorter than the observed timescale for the hard X-ray emission, and thus the observations
require either continuous acceleration and/or the presence of much more energetic electrons than
can be diagnosed with the spectral energy coverage and sensitivity of the Swift
detectors.  The first option is broadly consistent with hydrodynamic stellar flare models
which require continued heating \citep{reale1997}, hence continuous particle acceleration, assuming the two
processes are physically linked.  It is likely that both continuous acceleration and
a population of highly energetic electrons exist; radio observations of active stars
indicate the presence
of $\sim$ MeV electrons.  We expect that had radio observations of this flare been obtained,
the observations would have shown
a significant enhancement above typical levels.
Indeed, using the L$_{X}$--L$_{R}$ relationship of \citet{gb1993}, we would expect
a centimeter-wavelength radio luminosity of (4.5--45)$\times$10$^{17}$ erg s$^{-1}$ Hz$^{-1}$, or
0.2--2 Jy flux density, using the 0.8--10 keV X-ray luminosity during
Orbit 1 (L$_{X}$=7.7$\times$10$^{32}$ erg s$^{-1}$).
Although, in most solar flares, particle acceleration occurs only during the
impulsive phase, there is evidence for continued particle acceleration and chromospheric
evaporation during the gradual phase of some flares \citep{cliver1986,kai1986,sohocds}, which are
typically long-duration events.

\subsection{Occurrence Rate}
The Swift mission had been operating for roughly 9 months before this event was detected, observing
$\sim$60\% of the sky each day.  This suggests a rough frequency of occurrence for these superflares
of once every 5.4 months, or roughly once in 164 days.  
Other flare events from active binary systems with high luminosities
have been only sporadically reported in the literature.  If such events follow the usual power-law 
dependence of flare frequency with luminosity, 
then this is almost certainly due to the low probability of catching
such a flare during pointed observations lasting a few days or less.  
For the 66 days of exposure
accumulated in the {\it Ariel-V} observations of \citet{schwartz1981}, the rate of flares in excess of
$6 \times 10^{-10}$ erg cm$^{-2}$ s$^{-1}$ ($~10 - 30$ times the typical 
`quiescent' X-ray flux of $2 - 6 \times 10^{-11}$ erg cm$^{-2}$ s$^{-1}$ 
since determined for this system) in the 1.5-20 keV band was inferred to be $~11$ per
year. 

Dedicated surveys obviously will have a higher probability of detecting these events.  
\citet{ryle} have  a Ryle Telescope 15 GHz light curve of II Peg (their Fig.~13) which spans
14 months, albeit with some gaps, which shows one flare with an intensity greater than 200 mJy
($\approx$ 100 times the quiescent radio flux density) and 2 others $>$100 mJy.  
Based on several years' observation at centimeter wavelengths
with the Green Bank Interferometer reported in \citet{richards2003}, we can also estimate the occurrence
rate of radio superflares in active binary systems, using the flux data for HR~1099 and UX~Arietis
\footnote{Obtained from ftp://ftp.gb.nrao.edu/pub/fghigo/gbidata/gdata/gindex.html},
two analogous active binary systems to II~Peg containing a K subgiant and hotter companion (but where
the K subgiant, as in II~Peg, is presumed to dominate in the X-ray and radio emission).
Only for 0.2\% of the time did flare events in HR~1099 and UX~Ari reach radio luminosities in 
excess of  
100$\times$ the quiescent values, corresponding to a total of 9 flare events for these two
systems.  The average time between superflares is 136 days (165 days for 4 events on HR~1099 and
113 days for 5 events on UX~Ari); with the usual caveat of small number statistics,
the event rate for X-ray superflares and radio superflares appear to be broadly consistent, 
being roughly once every several months to a year.

\subsection{Contribution of Thermal and Nonthermal Energies}
A lower limit to the amount of thermal energy in the plasma can be obtained by 
determining the radiative flux from 
the thermal plasma over a large range of photon energies.  We did this in XSPEC through
the use of a ``dummy'' response  covering 0.01--200 keV.
For Orbits 1 and 2, we used the best-fit two-temperature plus Gaussian plus nonthermal model
parameters listed in Table~\ref{tbl:fits}, and removed the Gaussian and nonthermal model components.
A lower limit to the thermal energy can then be calculated as $4\pi d^{2}F \Delta t$,
where $F$ is the radiative flux from 0.01--200 keV (9.5$\times$10$^{-9}$ erg cm$^{-2}$ s$^{-1}$ for
Orbit 1 and 4.8$\times$10$^{-9}$ erg cm$^{-2}$ s$^{-1}$ for Orbit 2), and $\Delta t$ is the duration of
Orbit 1 and 2 (1535 and 2641 s, respectively).  In both cases, the radiative energy estimates
work out to $\approx$ 3$\times$10$^{36}$ erg, for a total of $\approx$ 6$\times$10$^{36}$ erg.
The total radiated energy of the hot plasma over all wavelengths can also be computed from the 
radiative loss function, using the radiative losses at the temperatures returned from spectral fitting
appropriate for the derived abundances, and multiplied by the emission measure at that temperature
and total duration.  We used the line and continuum emissivities in APEC to calculate
this function. 
The total radiated energy
calculated in this manner matches to within an order of magnitude the energy estimates
above; this is due partly to the wide wavelength range considered in the ``dummy'' response.
We note that the total radiated energy over all wavelengths from the few solar flares
in which changes in total solar irradiance have been made \citep{woods2004} indicate that
the total radiated flare energy 
can be larger than the radiated energy of the hot plasma by a
factor of up to ten.  Thus, a lower limit to the radiated energy for the flare in 
Orbits 1 and 2 combined
can be placed at $\sim$ 10$^{38}$ erg.
This is the amount of energy the plasma loses by radiation, and ignores the effect of
conductive energy losses (which will be important at high temperatures) as well as
bulk kinetic energy of the plasma or any energy loss due to expansion.

We can estimate the amount of conductive energy losses by using the equation for
conductive flux, \\
\begin{equation}
F_{\rm cond} = \kappa T^{5/2} \nabla T = \frac{\kappa T^{7/2}}{L} \;\;\; erg \;\;\; cm^{-2} \;\;\; s^{-1}\\
\end{equation}
where $\kappa$ is the Spitzer conductivity value ($=8.8\times10^{-7}$ erg cm$^{-1}$ s$^{-1}$ K$^{-7/2}$),
and we have approximated $\nabla T \sim T/L$ where L is a characteristic length along which 
conductive energy is lost.  The energy density per unit time can then be estimated as 
$F_{\rm cond}/L \sim \kappa T^{7/2}/L^{2}$. 
Using the relationship between volume, volume emission measure, and electron density ($VEM=n_{e}^{2}V$),
and the duration of each orbit, an estimate of conductive energy losses in Orbit 1 and 2 can be
estimated as \\
\begin{equation}
E_{\rm cond} = \frac{\kappa T^{7/2} VEM \Delta t}{L^{2}n_{e}^{2}} \;\; \; erg \; \; .
\end{equation} 
Using the temperature and volume emission measure\footnote{$\int n_{e}n_{H} dV$ tabulated in 
Table~\ref{tbl:fits} can be related to the volume emission measure $n_{e}^{2}V$ by using $n_{e}/n_{H}=$1.2.}
the conductive energy lost in Orbit 1 is $\sim$ 4$\times$10$^{43}$/(L$_{9}^{2}$n$_{10}^{2}$) erg,
while in Orbit 2 it is $\sim$6$\times$10$^{42}$/(L$_{9}^{2}$n$_{10}^{2}$) erg, for a length
in units of 10$^{9}$ cm and electron density in units of 10$^{10}$ cm$^{-3}$.
This is probably an upper limit on the amount of conducted energy, as evidence from solar flares
indicates \citep{jiang2006} that the classical expression for conductive flux may overestimate the total energy
decay rate.


We estimate the total energy represented by nonthermal electrons during
orbits 1 and 2 by multiplying the power output by the exposure time.  We
cannot diagnose variations 
in the acceleration of electrons on shorter timescales, due to signal-to-noise
limitations, so these calculations assume that electron acceleration occurs 
at a constant level during each orbit.
In Orbit 1 $F_{0}\times \Delta t=$10$^{40}$ erg, and in Orbit 2 $F_{0}\times \Delta t=$2 $\times$10$^{40}$ erg, where the low energy cutoff of 20 keV was used.
In order to match the thermal energy (here assumed to be dominated by conductive energy losses)
with the nonthermal energy input, we obtain a constrain on the product of electron density
and length scale, $L_{9}n_{10}=$60 for Orbit 1 and $L_{9}n_{10}=$20 for Orbit 2.
We have no independent constraints on either quantity.
We note that the reasonable agreement between the nonthermal energy and the upper and lower limits on
the thermal energy estimates add support to the nonthermal interpretation.

\subsection{Iron Fluorescence}
There is excess emission redward of the \ion{Fe}{25} and \ion{Fe}{26} features visible
in the XRT spectra of
Orbit 1 and 2, visible in the right-hand bank of panels in Figures~\ref{fig:spec1}
and ~\ref{fig:spec2}.
We attribute this to emission from the iron K$\alpha$ feature at 6.4 keV.
Both electron-impact ionization or photoionization mechanisms are capable of removing 
K shell electrons from neutral or near-neutral iron in the photosphere
and producing the 6.4 keV spectral line \citep{emslie1986},
although no solar K$\alpha$ line has been definitively identified with
electron-impact ionization \citep{parmar1984}.
We subdivided the spectra in orbits 1 and 2 to investigate the time evolution of the 6.4 keV feature
in more detail.
Figure~\ref{fig:feka} displays a close-up of the 5.5--8 keV region, and Table~\ref{tbl:feka}
lists the results of spectral fitting
for the five different time-resolved spectra.  

Previous observations of K$\alpha$ emission from stars other than the Sun
have been of pre-main sequence stars and have 
concluded that the fluoresced material is located in the circumstellar disk which is bathed
in the thermal hard X-ray continuum radiation emitted by flaring plasma.
Applying this interpretation to the event on II~Peg poses several problems, however.
The equivalent width of the 6708 \AA lithium resonance line and the lithium abundance are not consistent with 
a pre-main sequence evolutionary state, nor is the observed
C/N ratio \citep[see discussion in][]{berd1998}.  
The system is therefore generally considered to be an old-disk population star.
On the other hand, an
infrared
excess appears to have been detected in II~Peg \citep{lazaro1987,isoref} 
which is rather unexpected for a Pop I star.
Further, the column densities returned from X-ray spectral fitting are systematically higher
than previously determined interstellar values for this line of sight, suggesting
the possibility of additional absorption from a circumstellar disk.
However, we consider the column density returned from 
spectral fitting highly suspect, as we have excluded the energy
range below 0.8 keV due to systematics, and there could be additional unrecognized
systematics affecting the absorption column density.
Thus we consider it is more plausible that the 6.4 keV signature
arises from photospheric, not circumstellar, material, although we
cannot at the moment rule out a face-on circumstellar disk.


\subsubsection{Fluorescence}
If we attribute the formation mechanism to a photoionization fluorescence mechanism, then 
the continuum radiation above
the iron K edge, 7.11 keV, is the source of the photoionization producing the 6.4 keV feature.
The luminosity and equivalent width of the feature can be used to deduce the scale height of the
coronal emission above the photosphere, using equation 2 in \citet{coupref}\\
\begin{equation}
EW = \frac{L_{K\alpha}}{I(E_{K\alpha})} = \frac{\Delta \Omega}{4\pi}Y_{K\alpha} 
\frac{E_{K\alpha}}{I(E_{K\alpha})} \int n_{Fe}(s)ds \int_{\chi}^{\infty} 
\frac{I(E^{'})}{E^{'}} \sigma_{Fe}(E^{'}) dE^{'}
\end{equation}
where $\Delta \Omega$ is the solid angle subtended by the photosphere as seen by the flaring X-ray
source.
The continuum spectrum in this region is dominated by the hot thermal plasma, with a spectral shape
$I(E)\propto E^{-1}\exp^{-E/kT}$.
We use the fluorescence yield $Y_{K\alpha}$ of 0.342 \citep{bambynek1972}.
E$_{K\alpha}$=6.4 keV and $\chi$ is the iron K edge energy, 7.11 keV.  
The quantity $\sigma_{Fe}(E)$ is the photoelectric cross-section of iron; we use
$\sigma_{Fe}(E)=2\times10^{-20}(E/\chi)^{-3}$ cm$^{2}$ \citep{gullikson2001}.
We rewrite the integral involving $n_{Fe}(s)$ as $\int (A_{Fe}) n_{H} ds$,
where $A_{Fe}=n_{Fe}/n_{H}$.  
A comparison of coronal iron abundances \citep[here and in][]{chandraref}
and photospheric iron abundances does reveal
chemical fractionation occurring in II~Peg's atmosphere (in a sense opposite to that seen in the Sun), 
yet we cannot constrain such
spatial variations.  
The hydrogen column density in the chromosphere/photosphere is estimated using the column mass
density in the
atmospheric models of \citet{iipegchrmod}. Between the temperature minimum and
10$^{4}$ K, the column mass ranges from $\sim$0.001--2 g cm$^{-2}$, or
10$^{21}$--2$\times$10$^{24}$ cm$^{-2}$ with a mean molecular weight $\mu\sim$0.6. 
Since the photospheric iron abundance of II~Peg is higher than the coronal abundance by factors of 2--4,
and the hydrogen column density lower in the atmosphere exceeds that in the corona by orders
of magnitude,
we assume that the major contribution to this integral occurs in the photosphere
and express the integral as $A_{Fe, phot} \int n_{H} ds = A_{Fe,phot} N_{H}$, where
$A_{Fe,phot}\approx$0.4 A$_{Fe, solar}$ = 1.26$\times$10$^{-5}$, using the revised iron abundance of
\citet{gs1998},
and the above ranges of $N_{H}$.
In the solar case, the main contribution also occurs in the photosphere, due to the increasing
column mass as one proceeds from the corona to the photosphere; downward propagating
photons with E$>$7 keV are optically thin to photoelectric absorption and Compton scattering
\citep[see discussion in][]{parmar1984}, and the photospheric iron abundance being $\sim$one fourth that of the solar
coronal iron abundance does not affect where in the atmosphere the main contribution to fluorescence occurs
\footnote{Note that in the solar case, with an independent constraint on the height of the soft X-ray
source from e.g. X-ray imaging telescopes, the flux of the K$\alpha$ line can be used to deduce
the photospheric iron abundance \citep{bai1979}.}.

The solid angle $\Delta \Omega$ can be expressed in terms of a height using the
fraction of photons intercepted by the star, \citep{bai1979}, \\
\begin{equation}
\Delta \Omega(h) = 2\pi \left[ 1- \frac{\sqrt{h^{2}+2R_{\star}h}}{R_{\star}+h} \right]
\end{equation}
where $h$ is the scale height of the flaring X-ray source; we take
R$_{\star}$ to be the radius of the primary of the system, $\approx$ 3R$_{\odot}$ \citep{berd1998}.  Figure~\ref{fig:kalpha}
displays the relationship between height and column density for the measured
equivalent width and plasma temperature. There is an asympotic dependence
on column density, so that the minimum column densities which can reproduce the
equivalent widths are of order 10$^{24}$ cm$^{-2}$.  This would indicate, based 
on the atmospheric modelling of \citet{iipegchrmod}, that (1) the fluorescence occurred
deep in the atmosphere, near the temperature minimum region, and 
(2) the maximum flare scale height is 0.5R$_{\star}$.
There appears to be a discrepant behavior in the first sub-segment of Orbit 2, where the
equivalent width shows an anomalous value compared to Orbit 2b and Orbit 1c.
There are problems with this interpretation, however, due to the fact that 
at N$_{H}\sim1/\sigma_{T}\sim1.5\times10^{24}$cm$^{-2}$,
where $\sigma_{T}$ is the Thomson cross section, photons
start experiencing significant Compton scattering.
In contrast with the solar photosphere, in which the temperature minimum region
is reached with a lower column mass density \citep{val}, photons of energy 6.4 keV will also experience significant photoelectric absorption at
such high column densities, decreasing the efficiency of producing fluorescence emission in such
an environment. 
Thus we conclude that the fluorescence mechanism is probably not a valid
interpretation for the formation of the 6.4 keV line.  

The iron K$\alpha$ emission feature can have a great utility during stellar flares, if
further observations confirm the fluorescence mechanism. 
With an independent constraint on the flaring loop height, say from hydrodynamic 
flare modelling relating the loop height to the flare evolution in the T-VEM plane \citep{reale1997},
the equivalent width of the 6.4 keV feature can be used to constrain the photospheric $Fe/H$ value.
Line and continuum emission in the soft X-ray spectrum naturally constrain the coronal
$Fe/H$ value, allowing a simultaneous measurement of both with a single data set.  This would be
advantageous to studying chemical fractionation in stellar atmospheres, particularly as
photospheric abundances of active stars are notoriously difficult (due to e.g. fast rotation
and/or binarity).

\subsubsection{Collisional Ionization}
An alternative explanation raised for the production of the 6.4 keV feature seen in
solar flares is the collisional ionization of K shell electrons by a beam of nonthermal electrons
\citep{emslie1986}.  
We used equation 11 in \citet{emslie1986} to estimate the 6.4 keV line flux (in photons
cm$^{-2}$ s$^{-1}$) assuming the production mechanism is collisional ionization by a beam
of nonthermal electrons \\
\begin{equation}
\Phi = \frac{\omega \beta (n_{Fe}/n_{H})}{4\pi d^2 K \theta} \frac{\gamma-1}{\gamma} F_{0}(E_{low}) E_{low}^{\gamma-1} \times \int_{\chi}^{\infty} (E_{0}(E,N^{\star}))^{-\gamma} E Q_{I}(E) dE \; \;\;,
\end{equation}
where $\omega$ is the fluorescence yield of iron, $\beta$ is the branching ratio between
K$\alpha$ and K$\beta$, =0.882 \citep{bambynek1972}, $d$ is the distance to the object,
$K=2\pi e^{4} \Lambda/\theta^{2}$ with $e$ the electronic charge in e.s.u., $\Lambda$ the Coulomb
logarithm, $\theta$ the conversion from keV to erg, $=1.6\times$10$^{-9}$, $\gamma$ is the photon
spectral index (related to the spectral index of the electron distribution by $\gamma=
\delta-1$). The value $E_{low}$ is the low-energy cutoff of the electron distribution, fixed
to 20 keV in our spectral fitting.
The integral extends from the K$\alpha$ edge of 7.11 keV to infinity; $Q_{I}(E)$ is the 
collisional ionization cross section, obtained from the theory of \citet{am1958}.
$E_{0}$ is the initial energy of the electron at injection, and $E$ is its energy after collisional
(thick-target) encounters.  Specifically, $E_{0}$ satisfies the equation \\
\begin{equation}
E_{0}^{3} - 3KN^{\star}E_{0} = E^{3}
\end{equation}
where $N^{\star}$ is the column density in the K$\alpha$-emitting region.
The results of the thick-target bremsstrahlung spectral modelling during orbits 1 and 2
are used to estimate the 6.4 keV photon flux, as there is not enough signal to subdivide
the BAT spectra as was done above for the XRT spectra.  For $n_{Fe}/n_{H}$ we initially used
the photospheric iron abundance, $A_{Fe}$=1.26$\times$10$^{-5}$.  

We calculate the 6.4 keV line
flux relative to the underlying continuum flux (determined from the unfolded
spectrum in XSPEC) to determine the equivalent width of the feature.
Table~\ref{tbl:fits} lists the derived values for the results from thick-target spectral
modelling in Orbits 1 and 2.  
The right panel of Figure~\ref{fig:kalpha} displays the results for the thick-target
model in Orbits 1 and 2 as a function of N$^{\star}$.
The observed equivalent widths can be reproduced for values
of N$^{\star} \le$10$^{20}$ cm$^{-2}$.  
This mechanism is effective at lower column densities than a fluorescence mechanism implies,
and thus would take place higher in the atmosphere.
If the abundance
fractionation which is known to occur between the photosphere and corona in II~Peg is happening
at these column depths, the value of $n_{Fe}/n_{H}$ appropriate for the calculations 
must consequently be lower. The equivalent widths will then be lower, by a maximum of $\sim1/4$ the 
values in the Table~\ref{tbl:fits}, corresponding to the maximum iron depletion in the corona
as found by \citet{chandraref}.  
We conclude that the collisional ionization mechanism is to be preferred over the fluorescence mechanism.

\subsection{Loop Heights}
If we assume that the flare emission originated from an ensemble of coronal loops with
uniform cross section and roughly semi-circular mid-plane shape, then we can express
the observed
volume emission measure in terms of the loop height, density, number of loops,
and loop cross section.  Following Equation (5) of \citet{chandraref}, the loop height is \\
\begin{equation}
h=0.03 \left(\frac{N}{100} \right)^{-1/3} \left(\frac{\alpha}{0.1} \right)^{-2/3}
\left(\frac{VEM}{10^{53} cm^{-3}} \right)^{1/3} \left(\frac{n_{e}}{10^{11} cm^{-3}} \right)^{-2/3}
R_{\star} \;\; ,
\end{equation}
where the volume emission measure ($VEM$) is rewritten as the contribution from $N$ flaring loops, each with
height $h$ and aspect ratio $\alpha$, electron density n$_{e}$.
\citet{chandraref} deduced loop heights of 0.05 R$_{\star}$ under quiescent conditions, for $N=100$,
$\alpha=0.1$, $VEM=7.9\times10^{53}$ cm$^{-3}$ and $n_{e}\sim10^{11}$ cm$^{-3}$.
The peak emission measure determined here is $\approx$100 times larger than that
in \citet{chandraref}; applying the same
analysis to this superflare yields $h=0.3$R$_{\star}$, for the same density, for $N=100$.  
If instead only one loop is involved, $h$ could be as much as 1.3 R$_{\star}$.
An electron
density higher than the adopted value would result in more compact loops.
The fluorescence analysis of the 6.4 keV feature also yields a constraint on the solid angle extended by the
photosphere as seen by the continuum X-ray source, i.e. the scale height of the corona above
the photosphere.  The scale heights inferred in \S 4.5.1 are consistent with this simple 
scaling.

\section{Conclusions}
We have identified nonthermal emission as the most plausible mechanism
to explain the hard X-ray emission seen
during the rise and decay of a large stellar flare. 
This flare interestingly also displayed evidence for emission at 6.4 keV whose formation we attribute
to the mechanism of collisional ionization.
The increased sensitivity of the Swift BAT has
enabled a detection at much higher energies than had been possible with previous 
hard X-ray telescopes.
Nonthermal hard X-ray emission has enabled an investigation of the energetics of large stellar
flares, without complication from optical depth effects and source inhomogeneities.
The characteristics of this flare --- a rare, intense, transient event 
--- point to the value of triggered observations to study 
such flares; targeted observations would have had a very low likelihood of observing such an event
during the
short timescales over which the hard X-ray emission is produced.  Triggered observing modes
necessarily miss the preflare emissions, as the flux must pass a threshold value to warrant
a telescope slew.  Still, serendipitous science can be obtained, as this Swift-observed
flare demonstrates.  Current plans are to increase the trigger sensitivity threshhold of 
Swift by a factor of 4.5, thus enabling more such opportunities. 

Multi-wavelength observations of solar flares have revealed the dynamical response of the atmosphere
to these sudden intense inputs of energy.  
Future observations of stellar flares like the one discussed here can benefit from the
global array of telescopes which have been harnessed for gamma-ray burst studies. 
It will hopefully be possible to take advantage of
optical telescopes to reveal the response of the lower atmosphere and radio telescopes to
explore further the action of nonthermal particles.
Observations of such events will allow for a comparison of particle acceleration processes in active
stars and the Sun.

Support for this work was provided by NASA through Hubble Fellowship grant \# HF-01189.01 awarded
by the Space Telescope Science Institute, which is operated by the Association of Universities for
Research in Astronomy, Inc. for NASA, under contract NAS5-26555.  We are grateful to
Randall Smith for his assistance in implementing the thick-target bremsstrahlung code in XSPEC,
and in extending the APEC calculations to higher energies. RAO is also grateful for discussions
with Joel Allred about hard X-ray observations of solar flares.
The authors thank Brian Dennis, the referee, for a close reading of the paper and for
suggesting improvements to make the paper appeal to a wider audience.


\clearpage

\begin{deluxetable}{lllll}
\tablewidth{0pt}
\tablenum{1}
\tablecolumns{3}
\tablecaption{Trigger and Orbit 3 Spectral Fit Results \label{tbl:xspec}\tablenotemark{1}}
\tablehead{\colhead{Parameter} & \colhead{Trigger} & 
 \colhead{Orbit 3} \\
\colhead{} & \colhead{T-760:T+126\tablenotemark{2} }
& \colhead{T+10944:T+11479\tablenotemark{2} } 
}
\startdata
N$_{H}$ (10$^{22}$ cm$^{-2}$) & \ldots  &    0.095\\
        &\ldots   &  (0.07--0.12)\\
T$_{1}$ (10$^{6}$K)& \ldots    & 15\\
        & \ldots &  (14--16) \\
$(\int n_{e} n_{H} dV)_{1}$ (10$^{54}$ cm$^{-3}$) & \ldots &  15\\
         & \ldots  & (10--21)\\
T$_{2}$ (10$^{6}$K)& 177   & 61\\
        & (112--294) & (54--70)\\
$(\int n_{e} n_{H} dV)_{2}$(10$^{54}$ cm$^{-3}$) & 21. & 32.\\
        & (11.--39.) &  (30.--36.)\\
A\tablenotemark{a} & 1. & 0.16\\
        &  fixed &   (0.12--0.22)\\
$\chi^{2}$ (dof) & 23.4 (32) &  521 (514)\\
$f$(0.8--10 keV)\tablenotemark{b} (erg cm$^{-2}$ s$^{-1}$)& \ldots &  2.09$\times$10$^{-9}$ {\it (0.08)}\tablenotemark{c}\\
           & \ldots & 
(2.08--2.12)$\times$10$^{-9}$\\
$f$(10.-200. keV)\tablenotemark{b} (erg cm$^{-2}$ s$^{-1}$)& 1.2$\times$10$^{-9}${\it (0.04)}\tablenotemark{c} & \ldots\\
         & (0.8--1.3)$\times$10$^{-9}$ & \ldots \\
$\Delta t$(s) & 834 &  535 \\
\enddata
\tablenotetext{1}{Unless otherwise indicated, error ranges refer to 90\% confidence intervals.}
\tablenotetext{2}{Start and stop times of each time interval, in seconds since
the trigger time of 11:21:52 UT.  For the Trigger spectrum, these times refer to the BAT,
for Orbit 3 the time is the XRT.  }
\tablenotetext{a}{Scaled abundance relative to solar, using solar abundances of \citet{gs1998}.}
\tablenotetext{b}{Error ranges correspond to 68\% confidence intervals.}
\tablenotetext{c}{Number in italics refers to flux converted to luminosity and expressed as a fraction
of the bolometric luminosity.}
\end{deluxetable}

\begin{deluxetable}{lllll}
\tablewidth{0pt}
\tablenum{2}
\tablecolumns{5}
\tablecaption{Models Fit to XRT \& BAT Spectra in Orbits 1 and 2\label{tbl:fits}\tablenotemark{1}}
\tablehead{ \colhead{Parameters} & \colhead{1T+G\tablenotemark{2}} & \colhead{2T+G\tablenotemark{2}} & \colhead{3T+G\tablenotemark{2}} & \colhead{2T+G+NT\tablenotemark{2}} }
\startdata
\multicolumn{5}{c}{--- Orbit 1: T+126:T+1661\tablenotemark{a} ---} \\
\hline 
N$_{H}$ (10$^{22}$ cm$^{-2}$) & 0.062 &9.8e-2 & 0.1 & 0.1  \\
                        & (0.057--0.067) & (0.09-0.1) & (0.09--0.11) & (0.09--0.11) \\
A\tablenotemark{b} & 0.33 & 0.33  &0.38 & 0.34 \\
  & (0.28--0.38) & (0.27--0.38) & (0.31--0.49) & (0.29--0.39) \\
T$_{1}$ (10$^{6}$K)&139  & 16.5 & 15.4 & 15.3\\
             &(134--143) & (15.0--18.4) & (14.4--17.2) & (14.4--16.0) \\
($\int n_{e}n_{H} dV$)$_{1}$ (10$^{54}$ cm$^{-3}$) & 85 & 7.45 &5.3 & 5.8 \\
                       & (84--86) & (5.7--9.4) & (3.8--7.0) & (4.5--7.0)\\
N\tablenotemark{c} (10$^{-3}$ photons cm$^{-2}$ s$^{-1}$) & 2.4 & 2.5 & 2.6 & 2.7 \\
                       & (1.8--3.1) & (1.9--3.2) & (1.9--3.2) & (2--3.3) \\
EW\tablenotemark{d} (eV) & 45 & 48 & 47 & 51 \\
T$_{2}$ (10$^{6}$K)& \ldots & 152 & 118 &139  \\
                   & \ldots & (146--157) & (102--132) & (133--144) \\
($\int n_{e}n_{H} dV$)$_{2}$ (10$^{54}$ cm$^{-3}$) & \ldots & 83 & 65 & 83 \\
                   & \ldots & (82--84) & (47--77) & (83--84) \\
T$_{3}$ (10$^{6}$K)& \ldots & \ldots & 306 & \ldots \\
                & \ldots & \ldots & (224--580) & \ldots \\
$\int n_{e}n_{I} dV$ (10$^{54}$ cm$^{-3}$)& \ldots & \ldots & 20& \ldots \\
                    & \ldots & \ldots & (7--38)   & \ldots \\
$\delta$\tablenotemark{e} & \ldots & \ldots & \ldots& 2.8 \\
        & \ldots & \ldots & \ldots & (2.4--3.4) \\
F$_{0}$\tablenotemark{f} (10$^{36}$ erg s$^{-1}$) & \ldots & \ldots & \ldots & 8.5 \\
                                & \ldots & \ldots & \ldots & (7.6--28) \\
$\chi^{2}$ (dof) & 1065 (836) & 943 (834) & 918 (832) & 924 (832)\\
\hline 
\multicolumn{5}{c}{--- Orbit 2: T+4810:T+7451\tablenotemark{a} --- } \\
\hline
N$_{H}$ (10$^{22}$ cm$^{-2}$) & 0.011 & 0.066 & 0.067 & 0.066 \\
                           & (0.0055--0.016) & (0.057--0.075) & (0.058--0.077) & (0.058--0.075) \\
A\tablenotemark{b} & 0.31 & 0.26 &0.28 & 0.27 \\
   & (0.29--0.34) & (0.24-0.29) & (0.25--0.32) & (0.25--0.30) \\
T$_{1}$ (10$^{6}$K)& 67 & 15.5 & 15.3 & 15.4\\
               & (66--69) & (15.2--16.4) & (15.0--15.8) & (15.1--15.8) \\
$(\int n_{e}n_{H} dV)_{1}$ (10$^{54}$ cm$^{-3}$)& 58 & 13 &12 & 12 \\
                     & (57--59) & (12--15) & (10--14) & (11--14) \\
N\tablenotemark{c} (10$^{-3}$ photons cm$^{-2}$ s$^{-1}$) & 1.14 & 0.9 & 0.95 & 0.99 \\
                        &(0.84--1.6) & (0.56--1.3) & (0.6--1.32) & (0.65--1.3) \\
EW \tablenotemark{d} (eV) & 42 & 33 & 34 & 34 \\
T$_{2}$ (10$^{6}$K)& \ldots & 86 & 74 & 83 \\
                  & \ldots & (83--90) & (67--81) & (80--86) \\
$(\int n_{e}n_{H} dV)_{2}$ (10$^{54}$ cm$^{-3}$)& \ldots & 51 & 47 & 51.4 \\
                   & \ldots & (50--52) & (41--50) & (50.7--51.5) \\
T$_{3}$ (10$^{6}$K)& \ldots & \ldots & 302 & \ldots \\
                  & \ldots & \ldots & (191--1000) & \ldots \\
$\int n_{e}n_{I} dV$ (10$^{54}$ cm$^{-3}$)& \ldots & \ldots & 6.3& \ldots \\
           & \ldots & \ldots & (2.5--12) & \ldots \\
$\delta$\tablenotemark{e} & \ldots & \ldots & \ldots& 3.1 \\
         & \ldots & \ldots & \ldots & (2.2--4.1) \\
F$_{0}$\tablenotemark{f} (10$^{36}$ erg s$^{-1}$) & \ldots & \ldots & \ldots & 8.6 \\
                                 & \ldots & \ldots & \ldots & (6.0--26) \\
$\chi^{2}$ (dof) & 1460 (814) & 952 (812) & 932 (810) & 931 (810)\\
\enddata
\tablenotetext{1}{Uncertainty ranges refer to 90\% confidence intervals.}
\tablenotetext{2}{Key to model combinations:$n$T= $n$ thermal components, G=Gaussian
at 6.4 keV, NT=nonthermal thick target bremsstrahlung model.  See text for details.}
\tablenotetext{a}{Start and stop times of each orbit, in seconds since the trigger time
of 11:21:52 UT.}
\tablenotetext{b}{Abundance of thermal plasma as a multiple of the
solar photospheric metal abundance given in \citet{gs1998}.}
\tablenotetext{c}{Flux emitted in the 6.4 keV line.}
\tablenotetext{d}{Equivalent width of the 6.4 keV line.}
\tablenotetext{e}{Power-law index of accelerated electron spectrum; see text for
details.}
\tablenotetext{f}{Power in accelerated electrons from thick target bremsstrahlung modelling
above a cutoff energy of 20 keV.}
\end{deluxetable}


\begin{deluxetable}{llllll}
\tablewidth{0pt}
\tablenum{3}
\rotate
\tablecolumns{6}
\tablecaption{6.4 keV Feature\label{tbl:feka}\tablenotemark{1}}
\tablehead{ \colhead{Parameter} & \colhead{Orbit 1a} & \colhead{Orbit 1b}
&\colhead{Orbit 1c} &\colhead{Orbit 2a}& \colhead{Orbit 2b} \\
\colhead{} & \colhead{T+144:T+650\tablenotemark{2}} & \colhead{T+650:T+1156\tablenotemark{2}} & \colhead{T+1156:T+1662\tablenotemark{2}} 
& \colhead{T+4902:T+6174\tablenotemark{2}} & \colhead{T+6174:T+7447\tablenotemark{2}}
}
\startdata
N$_{H}$ (10$^{22}$ cm$^{-2}$)& 0.086 & 0.1 & 0.11 & 0.075 & 0.048 \\
               & (0.07--0.10) & (0.087--0.12) & (0.09--0.12) & (0.065--0.092) & (0.035--0.061) \\
 A\tablenotemark{a} &0.47 &0.35 &0.26 & 0.28&0.29 \\
   & (0.36--0.58) & (0.26--0.44) & (0.17--0.35) & (0.23--0.32) & (0.24--0.33) \\
T$_{1}$ (10$^{6}$K) & 15.1 & 15.3 & 15.8 & 16.2 & 15.4 \\
                & (12.76--18.56) & (13.57--18.7) & (14.6--18.9) & (15.1--17.4) & (14.8--16.1) \\
($\int n_{e}n_{H}dV$)$_{1}$ (10$^{54}$ cm$^{-3}$) & 3.0 & 6.0 & 11.1 & 15.1 & 10.4 \\
               & (1.7--5.1) & (3.8--8.7) & (7.4--17.7) & (12.3--18.5) & (8.7--12.8) \\
T$_{2}$ (10$^{6}$K)& 151  & 142  & 140  &88  & 86 \\
        & (138--173)& (130--155)& (126--169) & (82--95) & (81--91) \\
($\int n_{e}n_{H}dV$)$_{2}$(10$^{55}$ cm$^{-3}$)& 6.8  & 8.7 & 9.3 & 5.3  & 4.9   \\
      & (6.6--6.9) & (8.5-8.8) & (9.1--9.4) & (5.1--5.5) & (4.8--5.0) \\
Flux in 6.4 keV feature & 2.5  & 2.7 & 2.8  & 0.65  & 1.68 \\
   (10$^{-3}$ photons cm$^{-2}$ s$^{-1}$)  & (1.4--3.6) & (1.5--3.9) & (1.5--4.0) & (0.08--1.1) 
& (1.2--2.2) \\
EW (eV)\tablenotemark{b} & 55 & 48 & 47 & 18 & 61\\
$\chi^{2}$ (dof) & 526 (552) & 549 (538) & 592 (555) & 595 (579) & 621 (555) \\
\enddata
\tablenotetext{1}{Error ranges refer to 90\% confidence intervals.}
\tablenotetext{2}{Start and stop times in Orbits 1 and 2
during which XRT spectra were extracted, in seconds since
the trigger time of 11:21:52 UT. }
\tablenotetext{a}{Abundance of thermal plasma as a fraction of the solar photospheric metal
abundance of \citet{gs1998}.}
\tablenotetext{b}{Equivalent width of the 6.4 keV line.}
\end{deluxetable}


\clearpage

\clearpage
\begin{center}
\includegraphics[scale=0.3,angle=270]{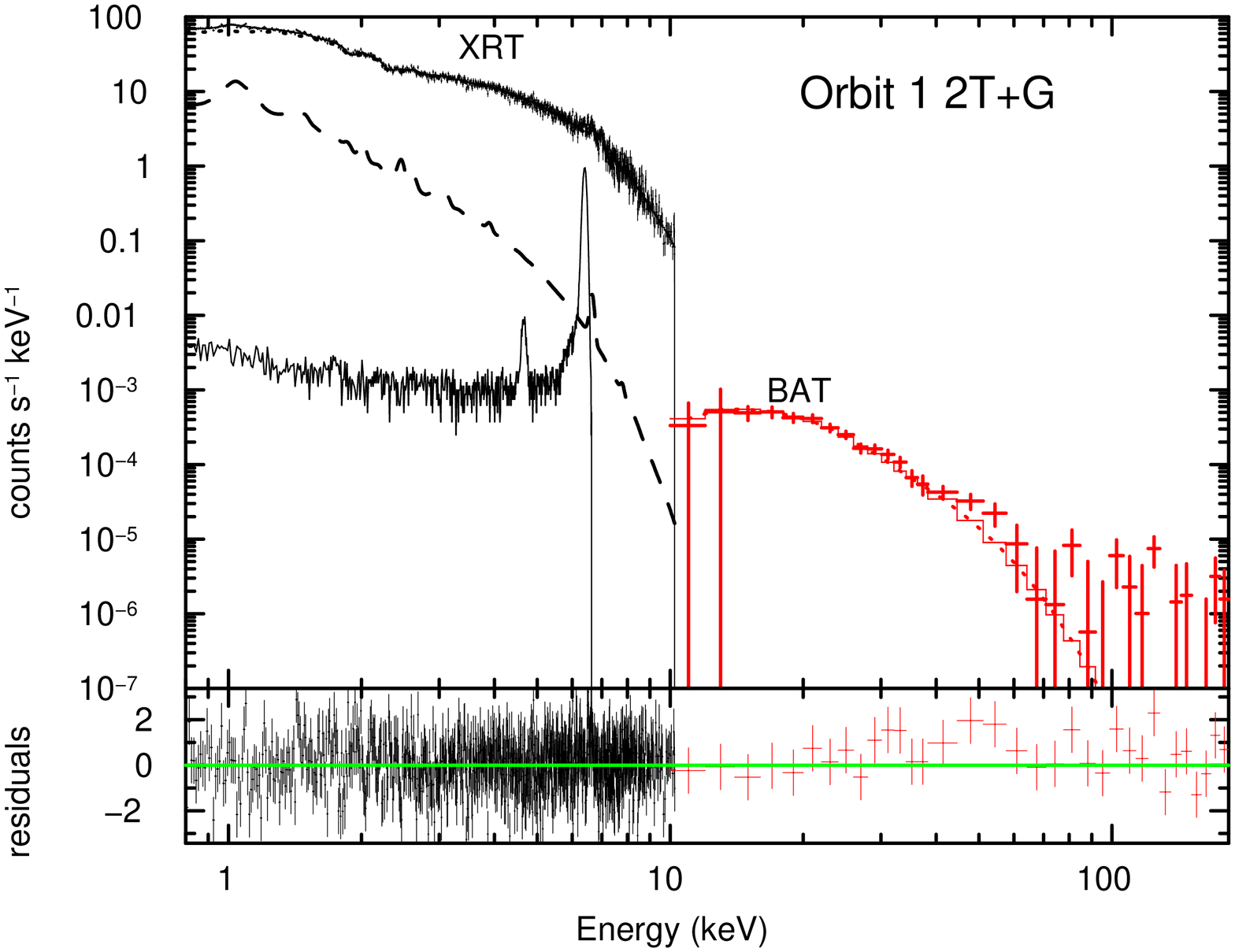}
\includegraphics[scale=0.3,angle=270]{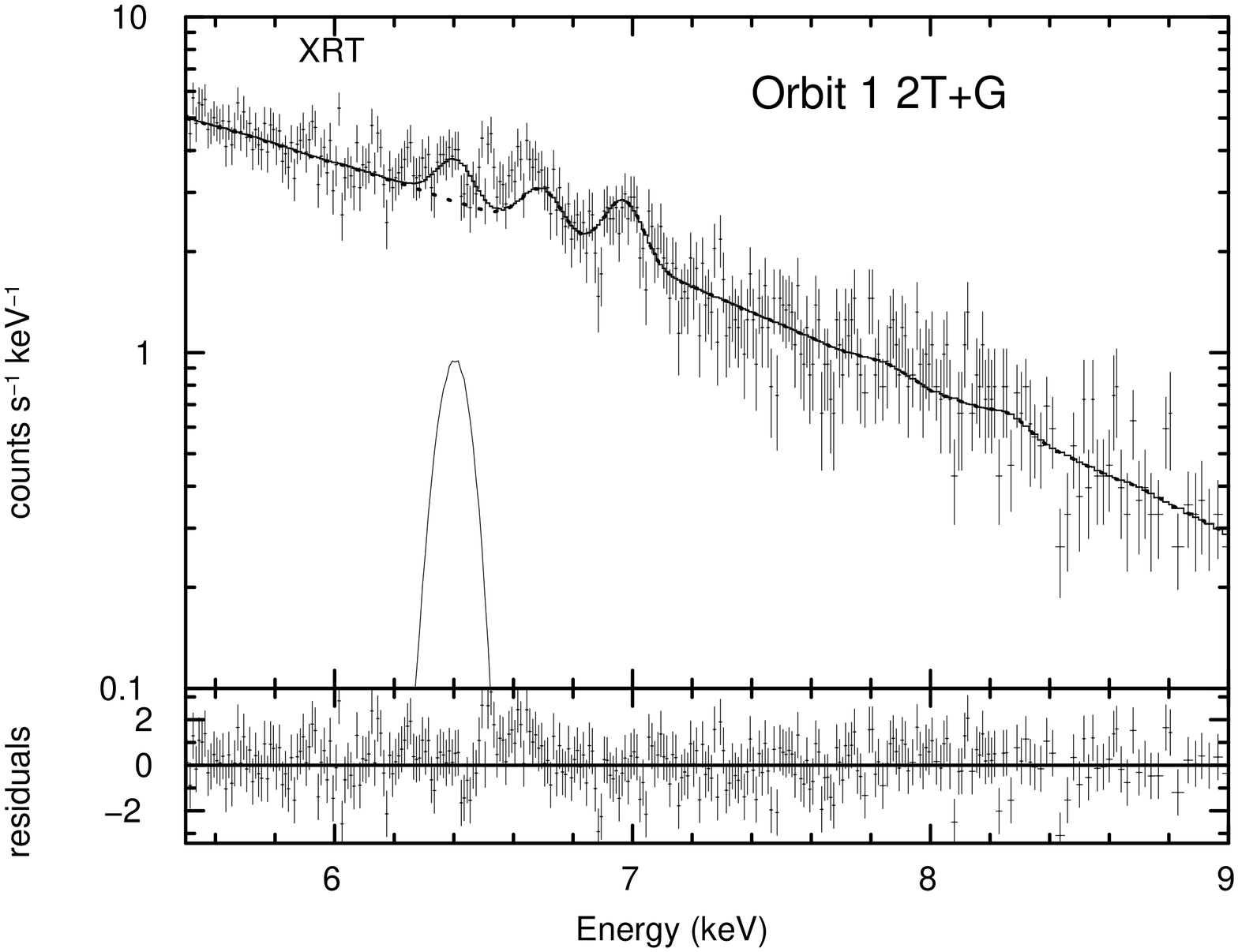}
\includegraphics[scale=0.3,angle=270]{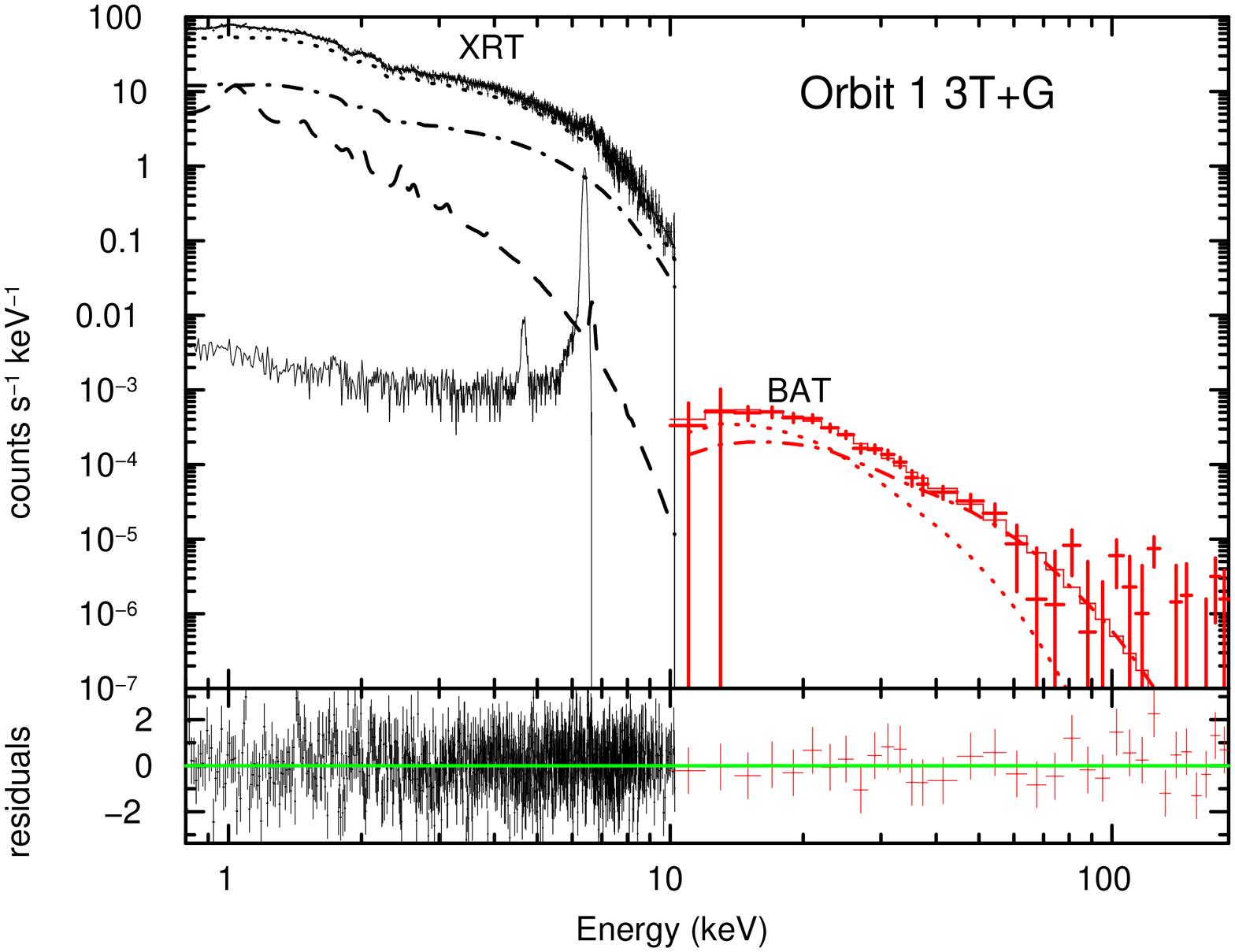}
\includegraphics[scale=0.3,angle=270]{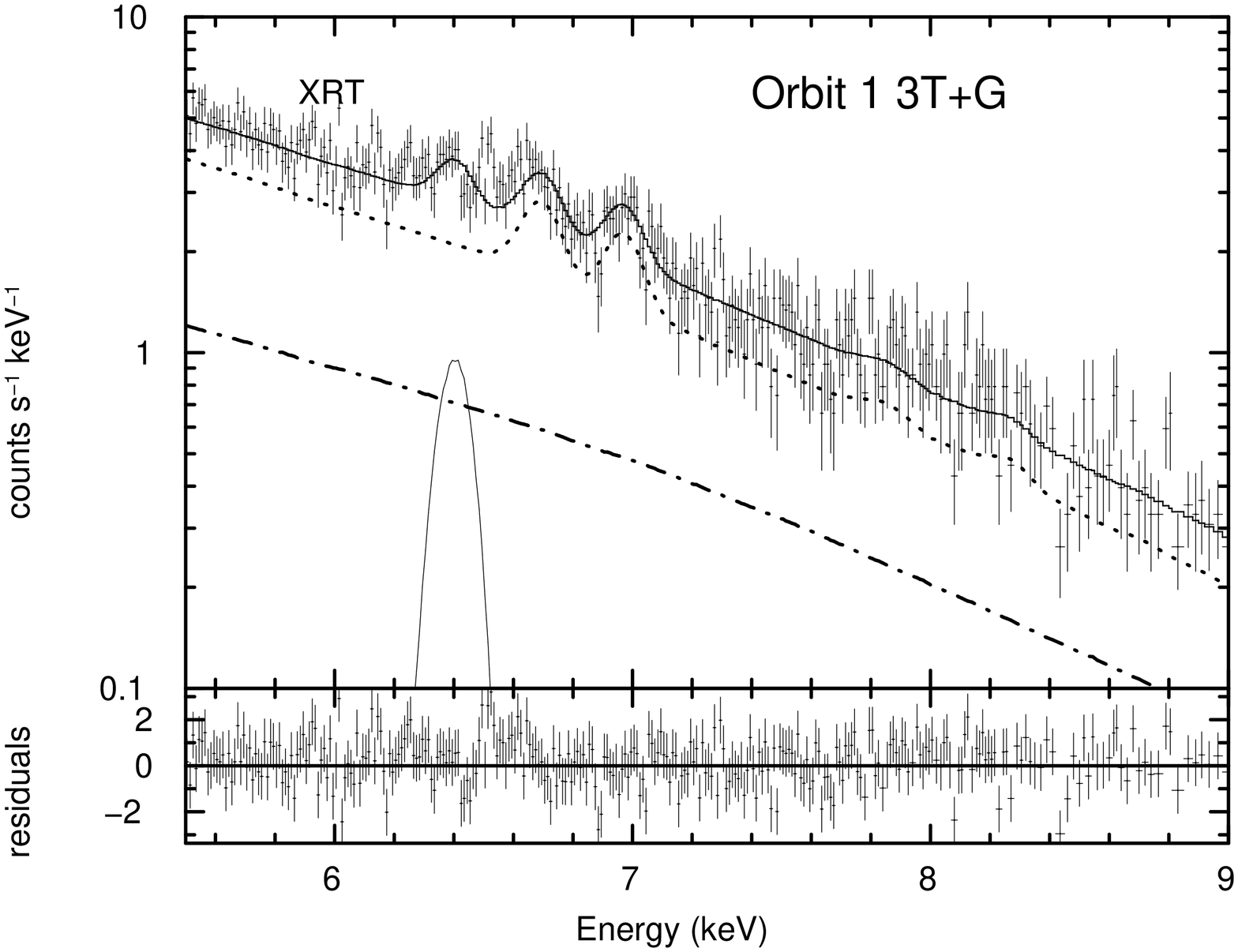}
\includegraphics[scale=0.3,angle=270]{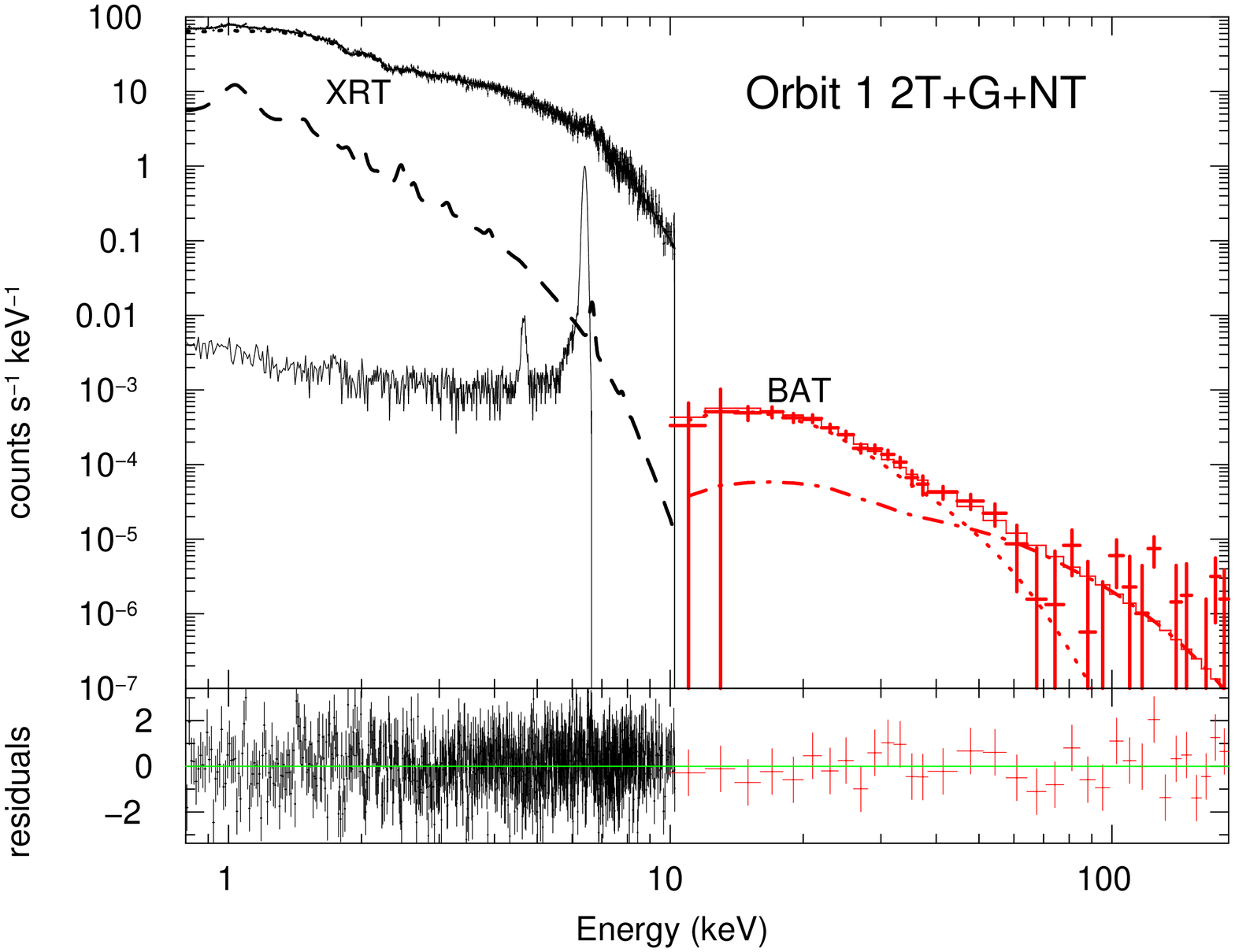}
\includegraphics[scale=0.3,angle=270]{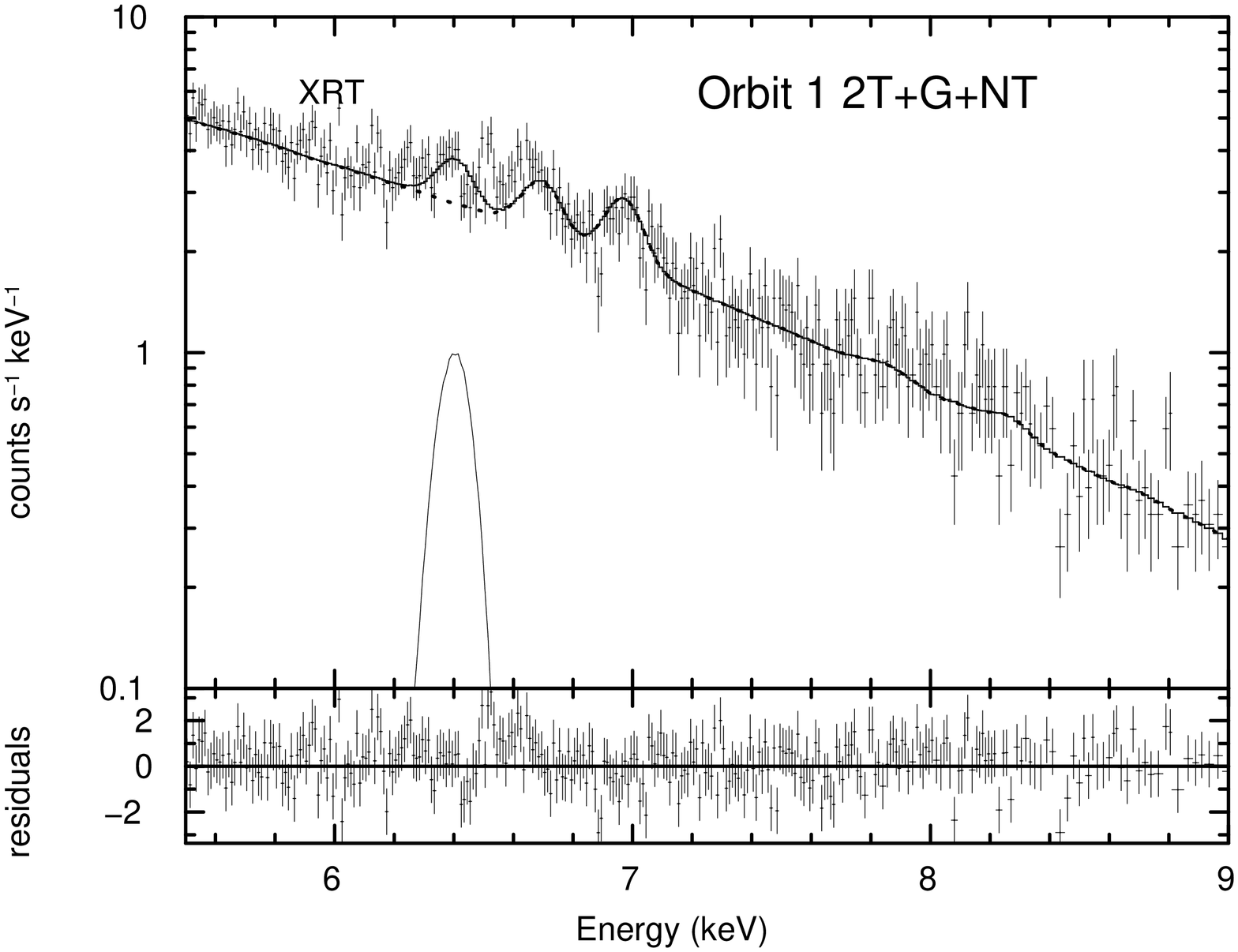}
\end{center}
\clearpage
\begin{figure}[h]
\figurenum{3}
\figcaption[orbit1_2tspec.ps]{Spectral fits to XRT and BAT spectra in Orbit 1.  
The top panel of each sub-plot displays data and model,  while the bottom panel
plots the residuals between data and model.  The residuals are in units of
$\sigma$ and are plotted with unity error bars.
In the upper panel of each sub-splot, crosses show spectral data points with errors.  
The histogram 
displays the final model with all components included.
The contribution of each additive component is also shown:
the first temperature component is delineated by a dashed line, the second
by a dotted line, and the Gaussian at 6.4 keV by the thin solid line.
The large residuals above 40 keV in the two temperature fit require the 
addition of another model component.
In models with either a third temperature component or a nonthermal component, this
additive component is indicated by a dashed-dotted line.
The left bank of sub-plots shows the data from 0.8--200 keV, while the right bank of sub-plots 
zooms in on the Fe~K region.  
The He- and H-like transitions of Iron confirm the
2nd temperature component in all cases.  
See Table~\ref{tbl:fits} for fit parameters.
\label{fig:spec1}}
\end{figure}
\clearpage
\begin{figure}[h]
\figurenum{4}
\begin{center}
\includegraphics[scale=0.3,angle=270]{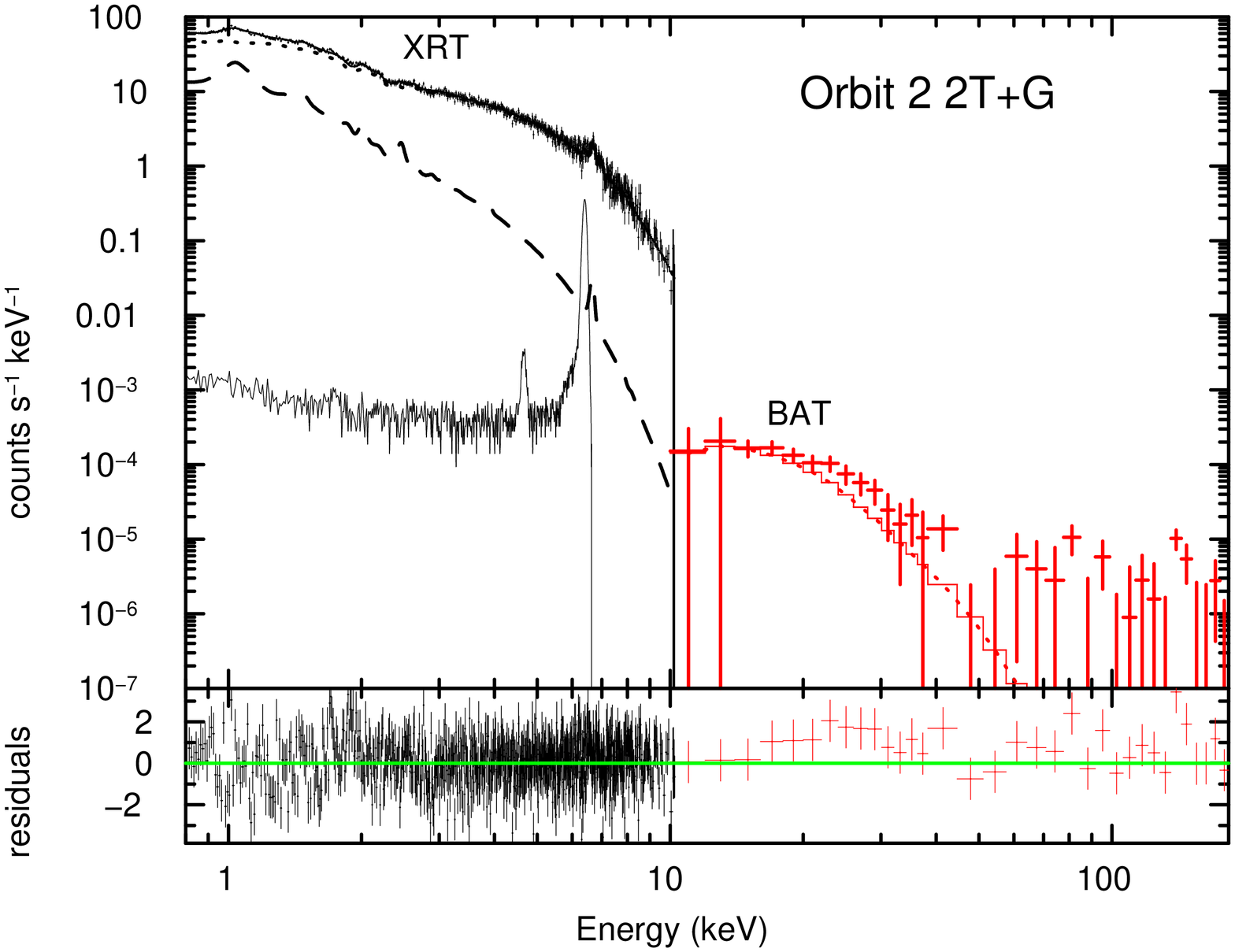}
\includegraphics[scale=0.3,angle=270]{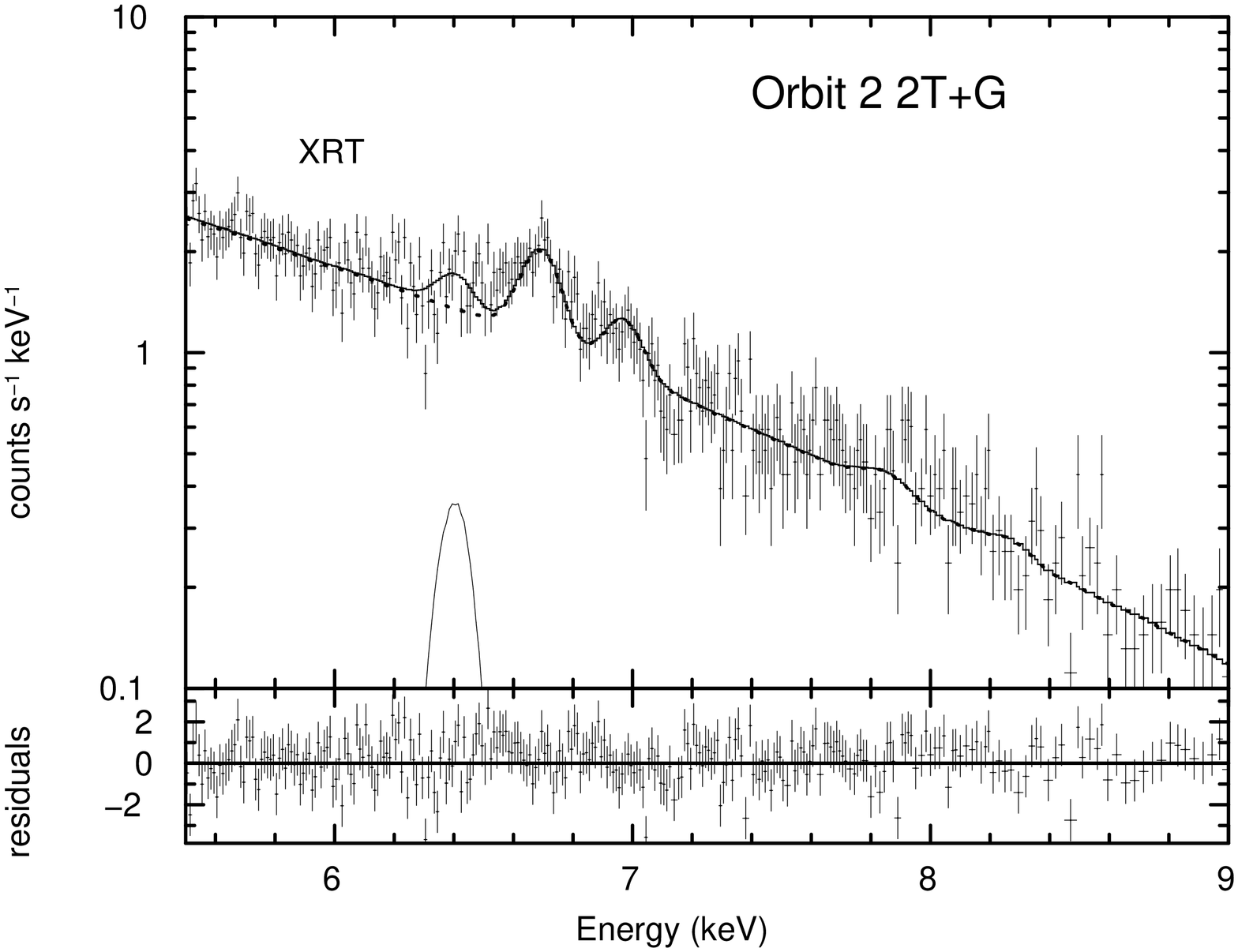}
\includegraphics[scale=0.3,angle=270]{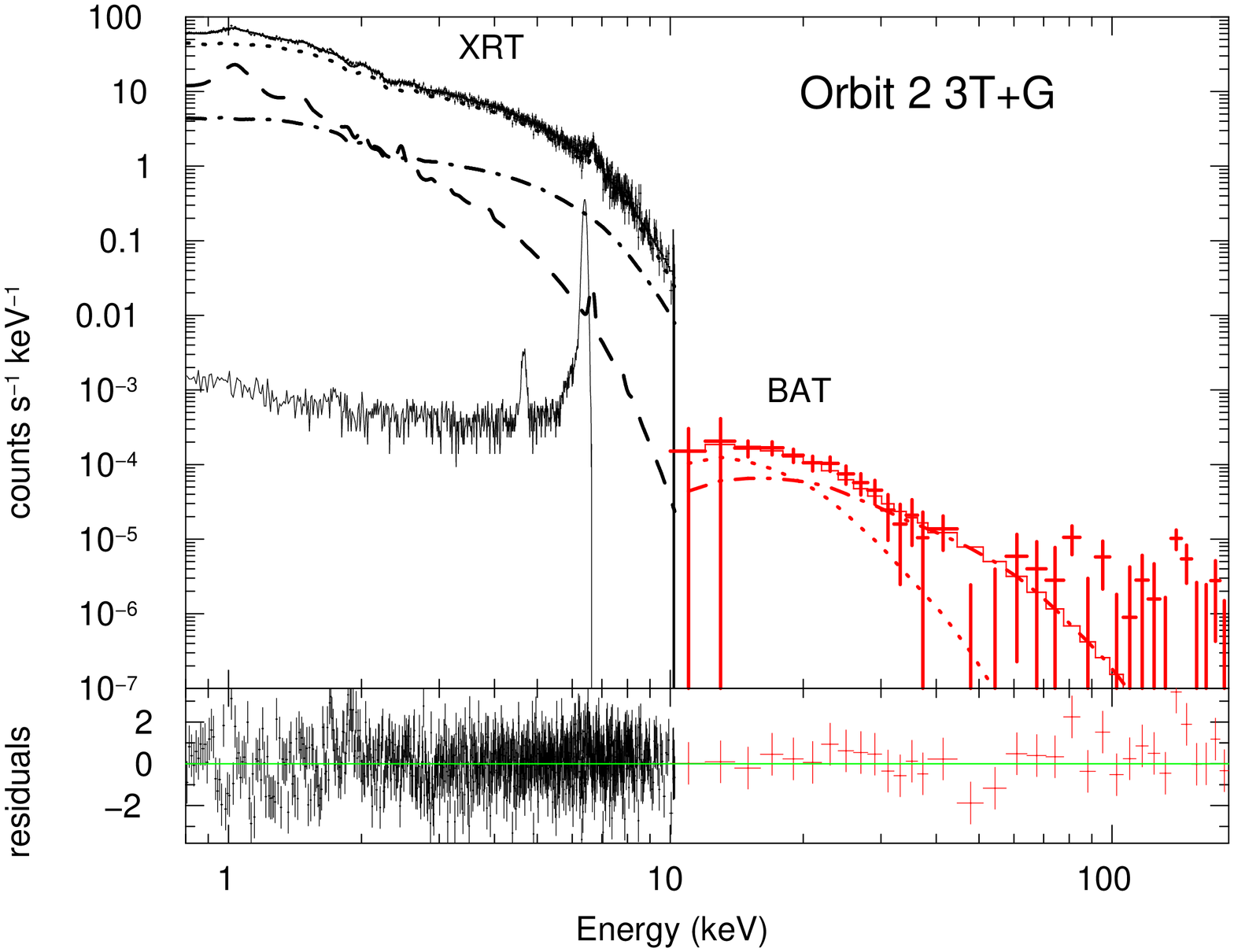}
\includegraphics[scale=0.3,angle=270]{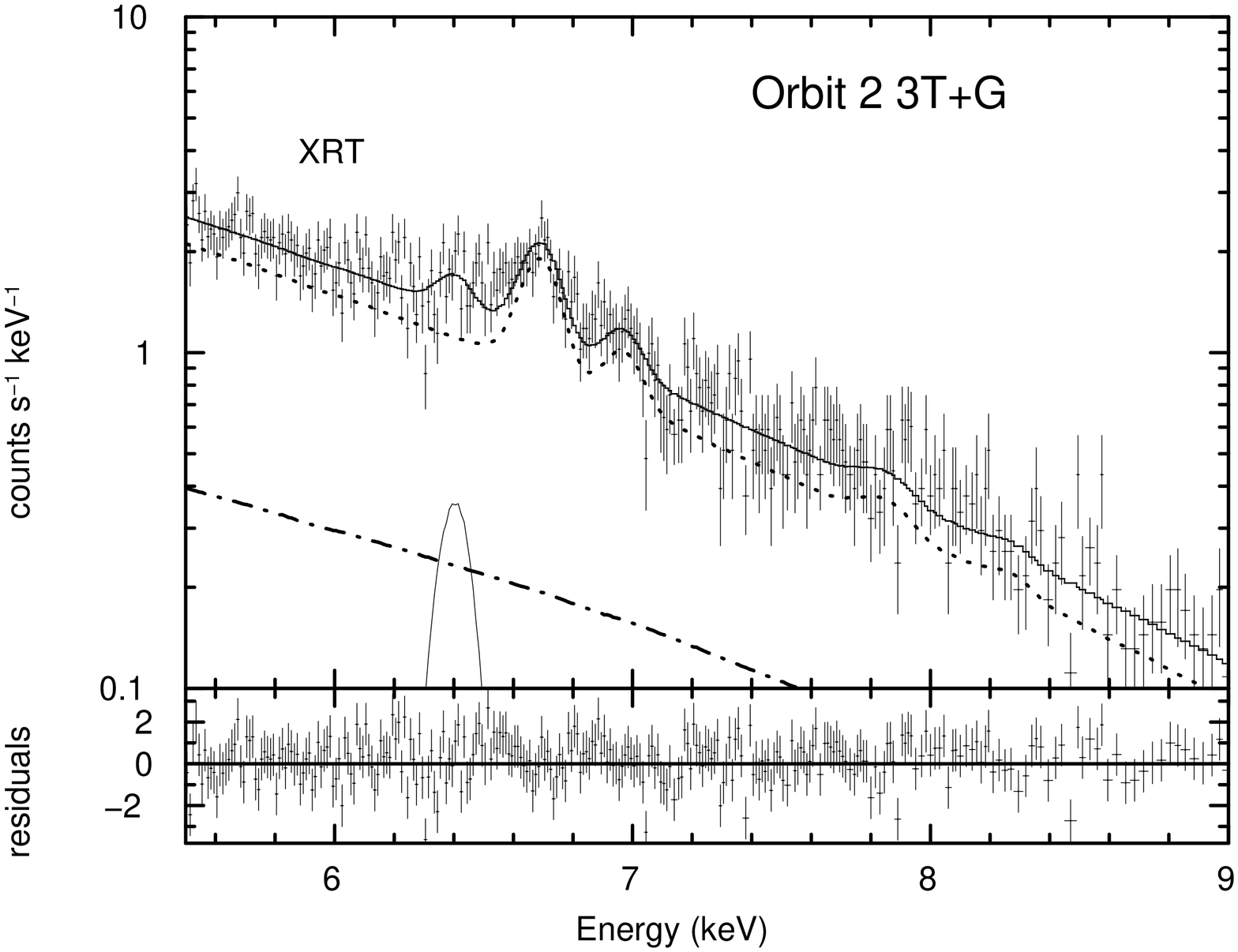}
\includegraphics[scale=0.3,angle=270]{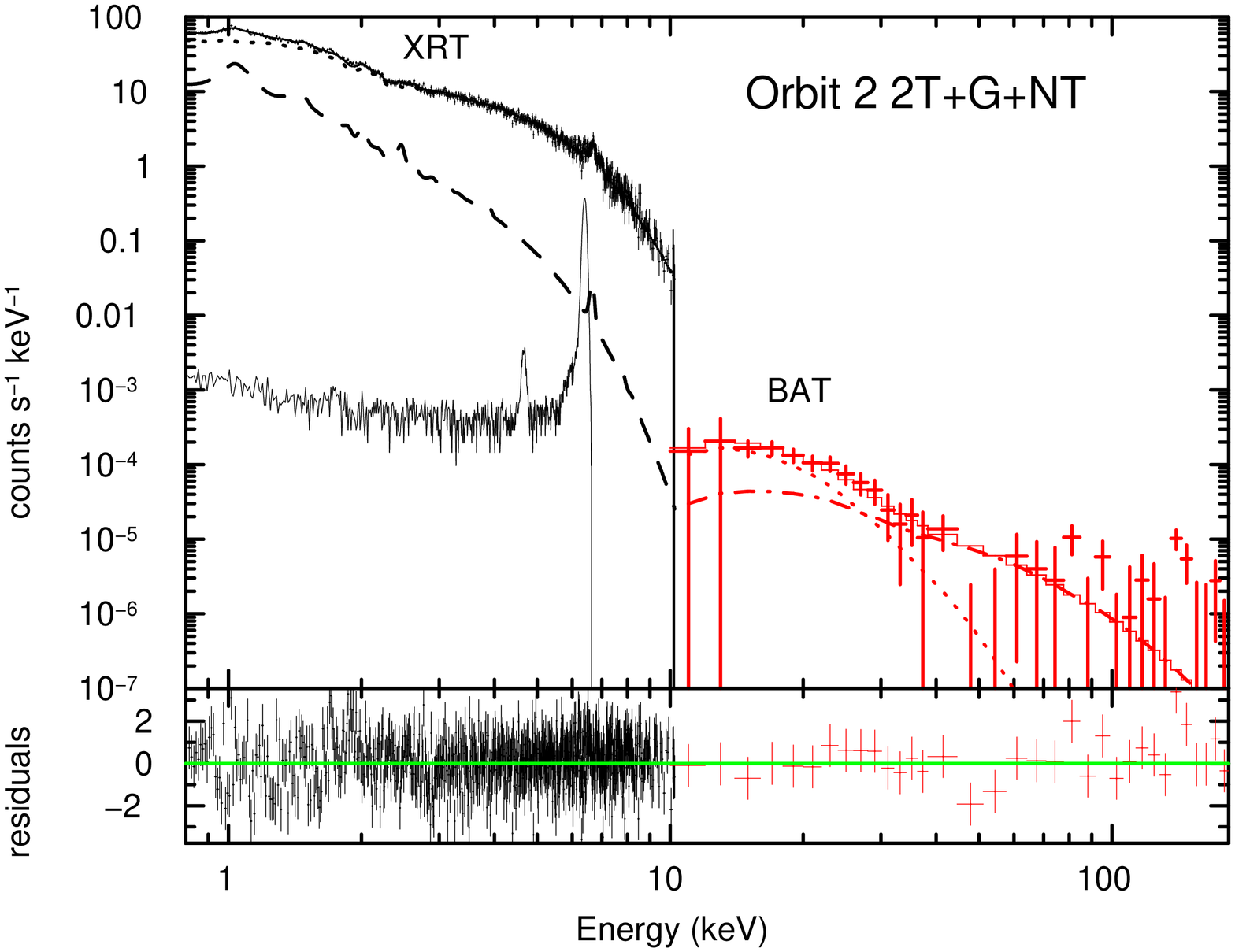}
\includegraphics[scale=0.3,angle=270]{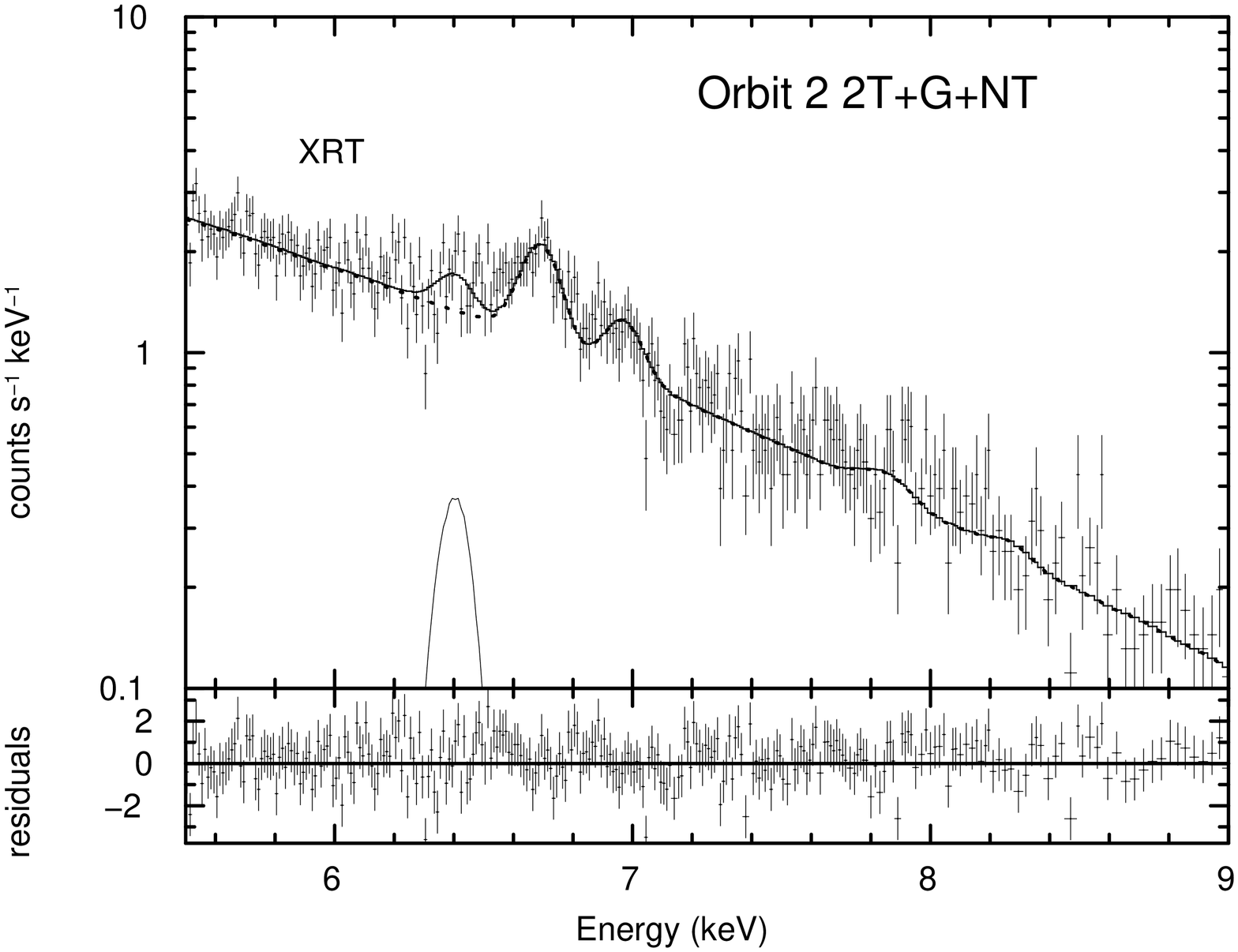}
\figcaption[]{Same as Figure~\ref{fig:spec1}, but for Orbit 2.  Note the 
residuals between data and model for the two temperature fit in the left-hand panel show a systematic
excess above about 20 keV, which can be fit by either a third temperature or a nonthermal
model. \label{fig:spec2}}
\end{center}
\end{figure}

\begin{figure}[h]
\begin{center}
\figurenum{5}
\includegraphics[scale=0.5]{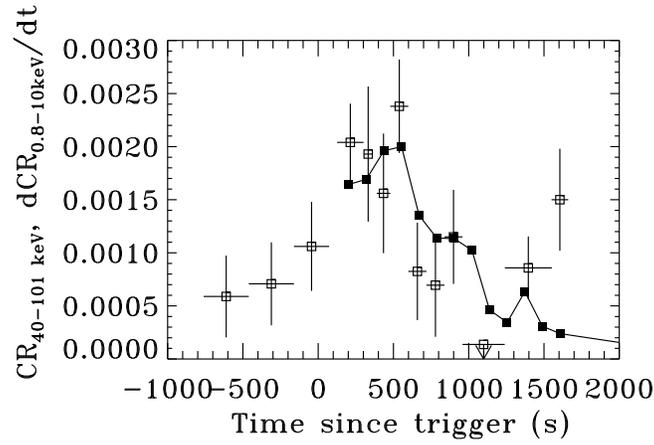}
\figcaption[]{Plot of 40--101 keV BAT light curve (heavy squares with error bars), 
along with scaled derivative of 0.8--10 keV XRT light curve (filled squares connected by a line) for the first 2500 seconds of the flare.
The XRT light curve has been rebinned to a coarser time sampling than original data.
The correlation for roughly one thousand seconds after the trigger may be indicative of the 
Neupert effect relating energy deposition in nonthermal particles and subsequent plasma heating.
\label{fig:neupert}}
\end{center}
\end{figure}

\begin{figure}[h]
\begin{center}
\figurenum{6}
\includegraphics[angle=270,scale=0.3]{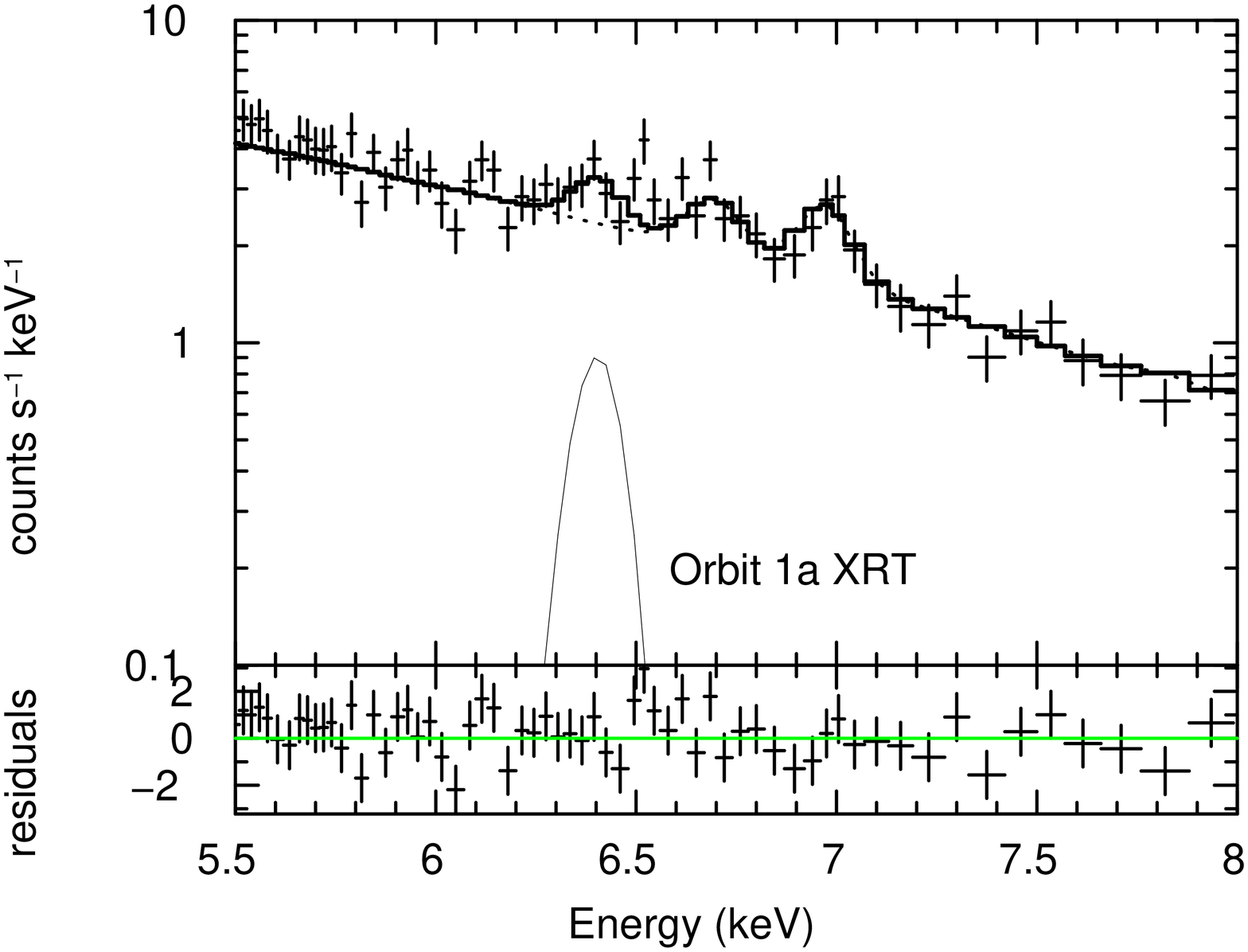}
\includegraphics[angle=270,scale=0.3]{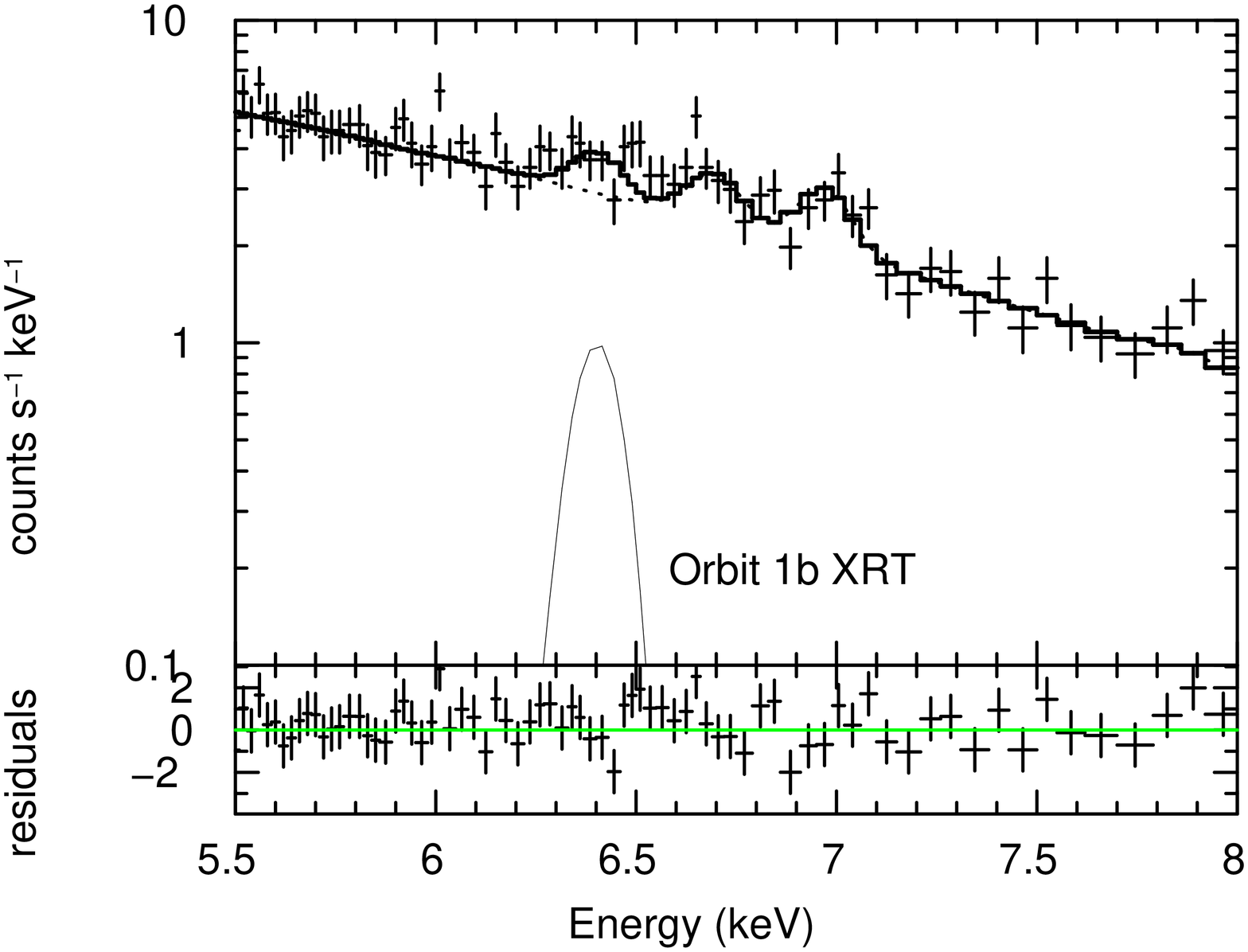}
\includegraphics[angle=270,scale=0.3]{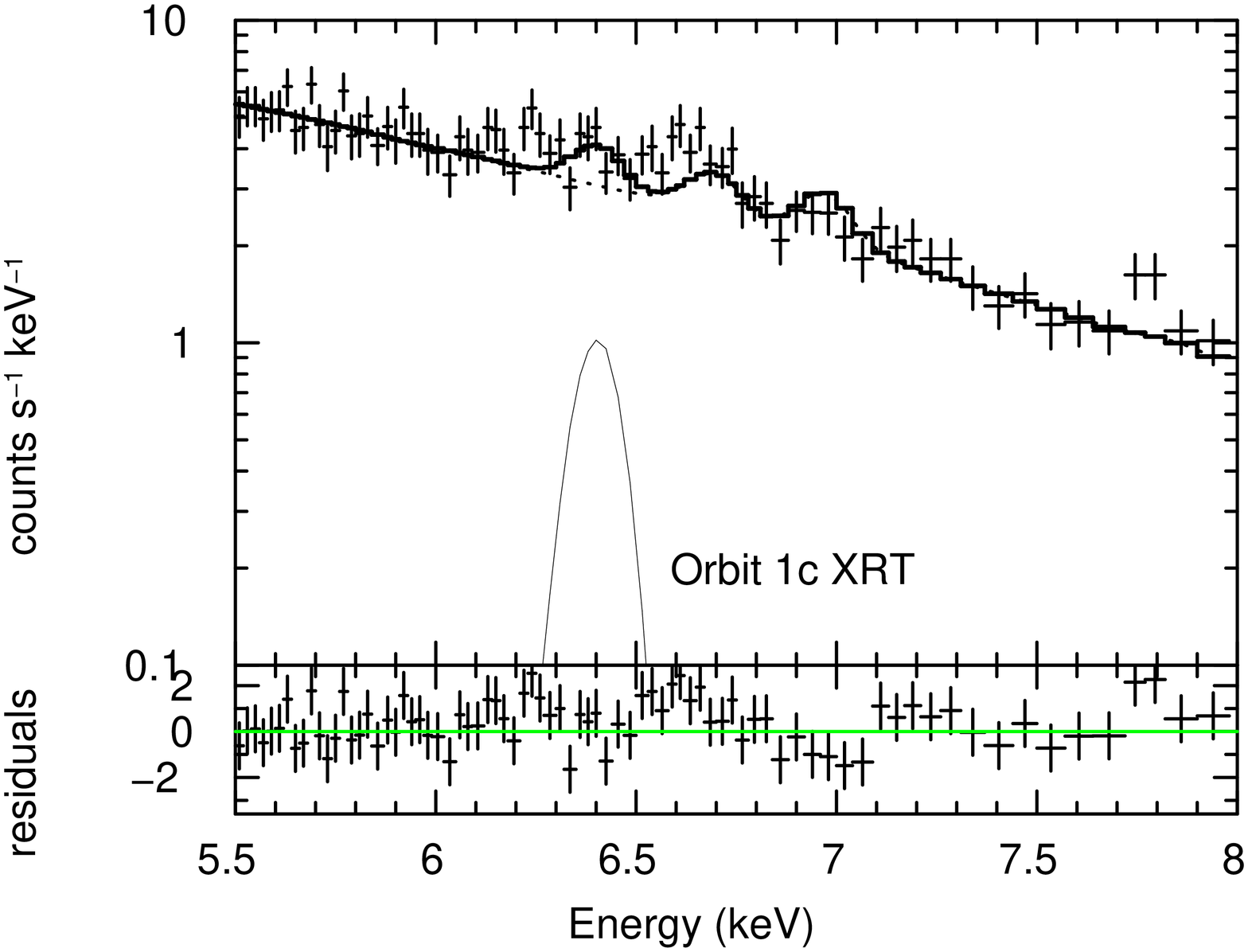}
\includegraphics[angle=270,scale=0.3]{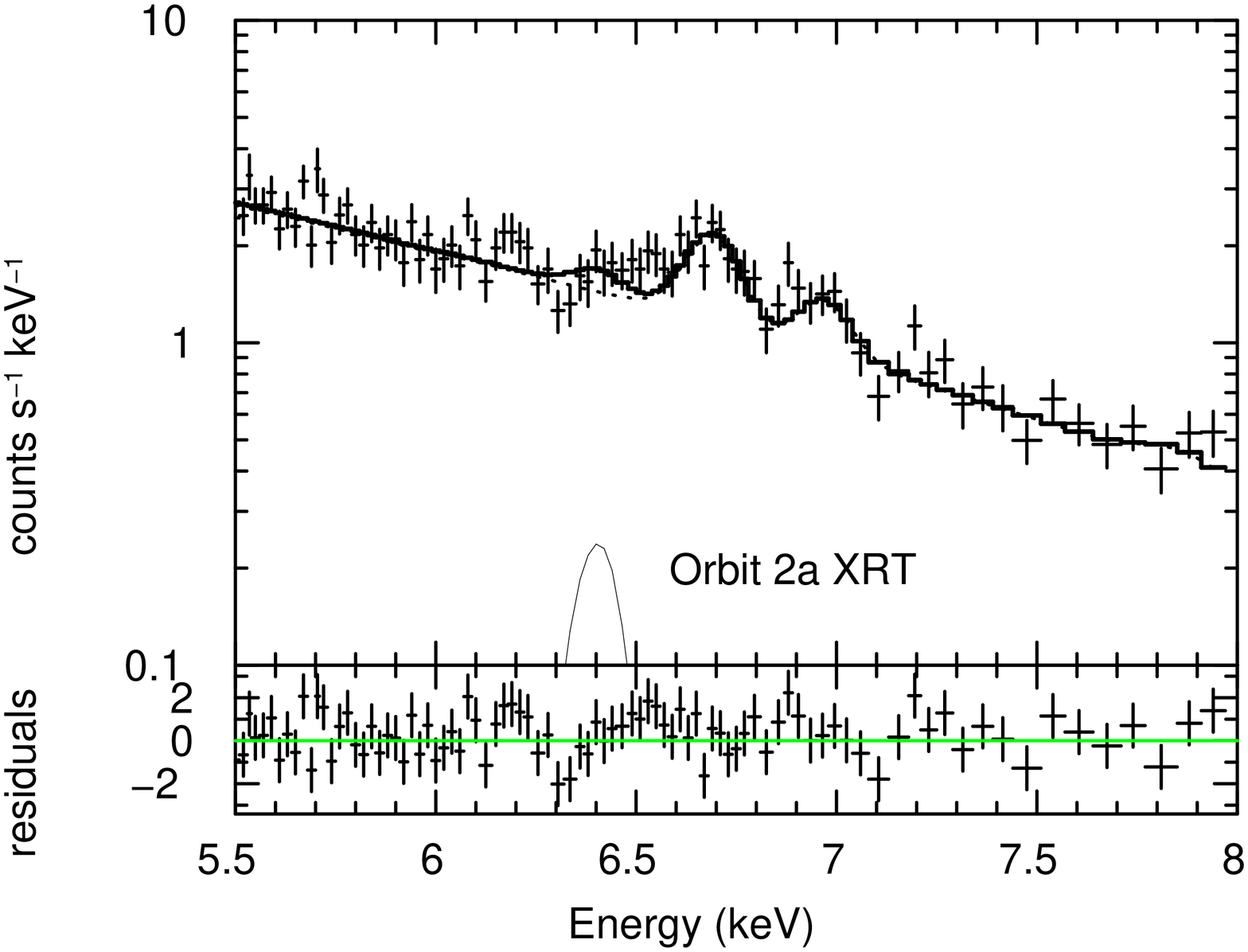}
\includegraphics[angle=270,scale=0.3]{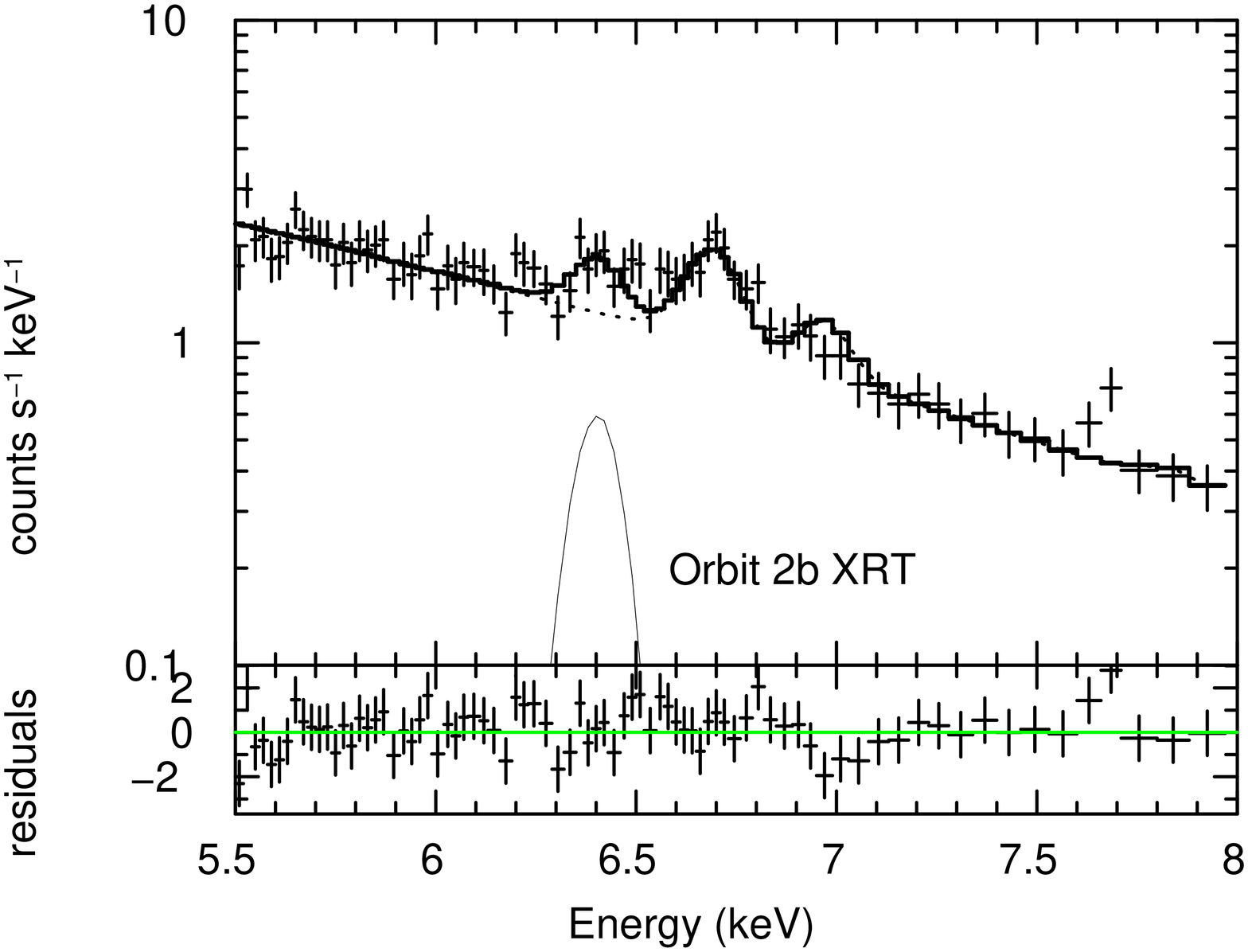}
\figcaption[]{ Intra-orbit variation of the spectral region around the
Fe K$\alpha$ 6.4 keV feature, for three time
slices in Orbit 1 and two in Orbit 2.  
Time intervals are listed in Table~\ref{tbl:feka}
Data points and errors are shown as crosses.
The 6.4 keV component is drawn with a solid line, and the dotted 
line delineates the emission from the APEC model; the thick histogram
describes the total model emission over this energy interval.
The prominent features at 6.7 and 6.9 keV are the He- and H-like
transitions of iron, respectively.
``Resid'' refers to residuals between model and data in each energy bin, in 
units of $\sigma$.
See Table~\ref{tbl:feka} for spectral fit results. 
\label{fig:feka}}
\end{center}
\end{figure}

\begin{figure}[h]
\begin{center}
\figurenum{7}
\includegraphics[scale=0.35]{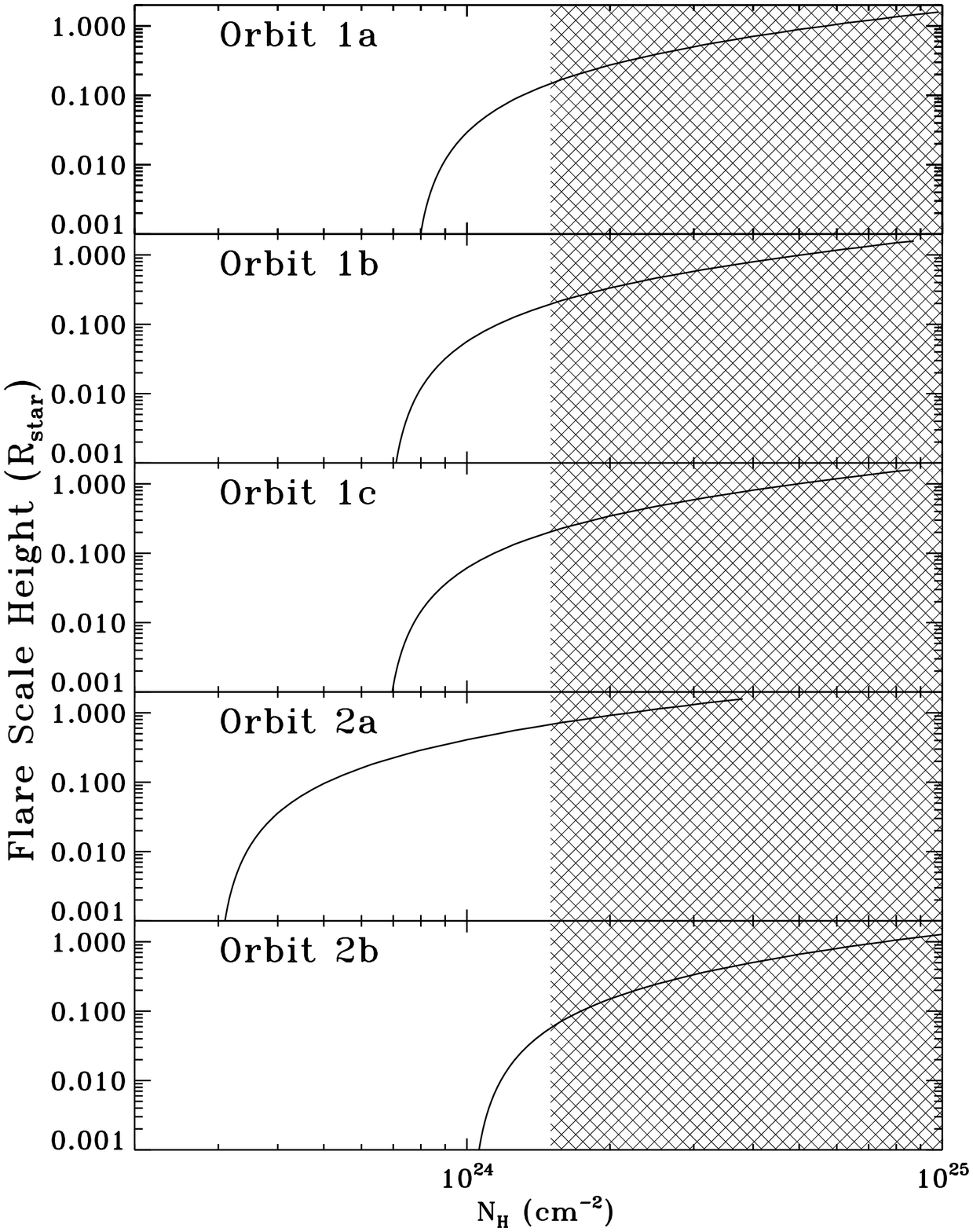}
\includegraphics[scale=0.4,width=3in,height=3.8in]{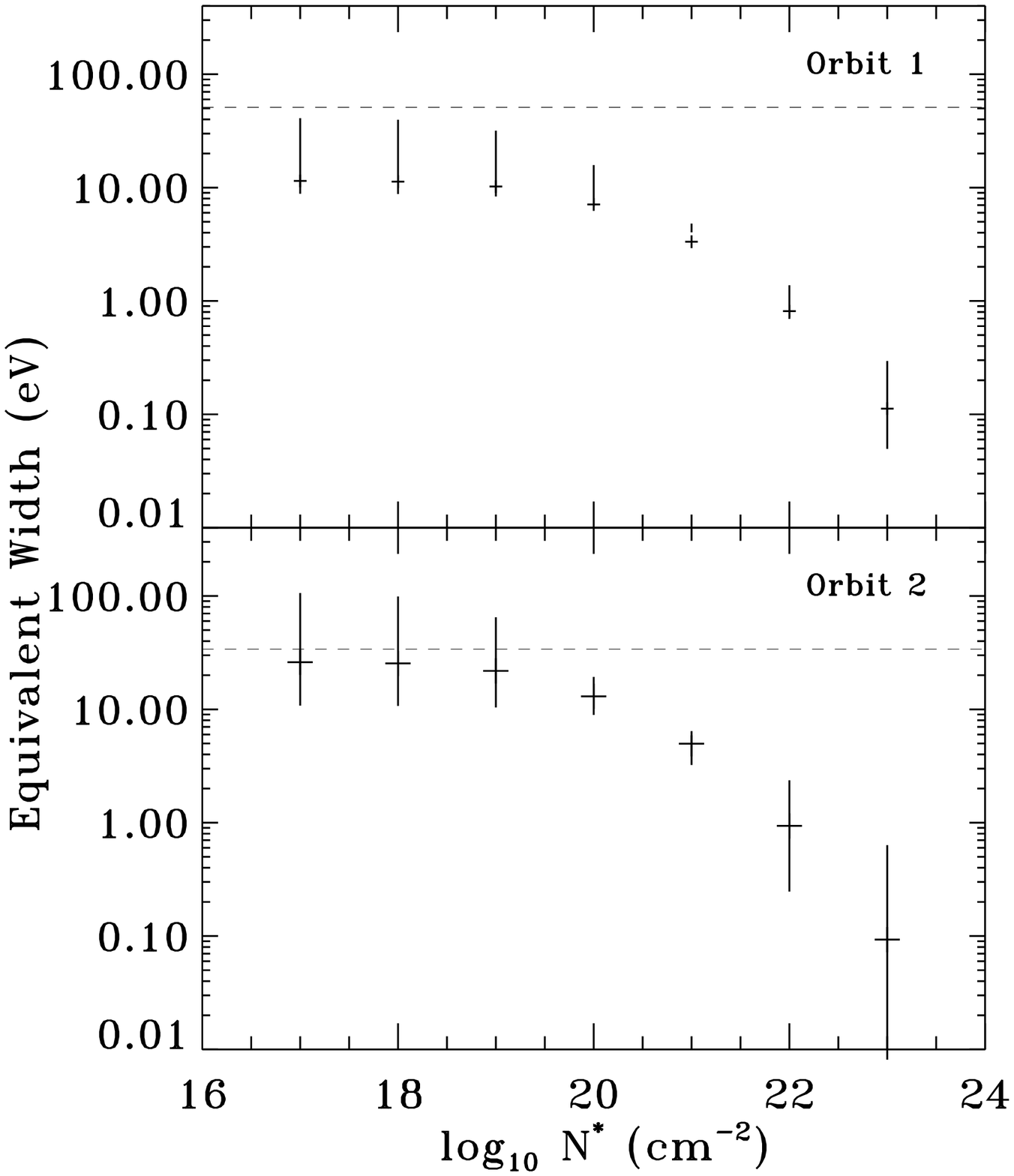}
\figcaption[]{ {\it (left)} Relationship between flare scale height and column density in the
lower atmosphere of II~Peg, using the plasma temperature 
and equivalent widths appropriate to each time slice (values listed in Table~\ref{tbl:feka}), 
for five time intervals in orbits 1
and 2.  Above column densities of $N_{H}\sim$1.5$\times$10$^{24}$ cm$^{-2}$, significant Compton
scattering will occur and 6.4 keV emission will be reduced or eliminated;
this region is indicated by the cross-hatching.
See \S 4.5.1 for discussion. 
{\it (right)} Predicted equivalent width as a function of $N^{\star}$, the atmospheric
column density where K$\alpha$ emission originates, for collisional
ionization formation of the 6.4 keV K$\alpha$ line.  Top panel shows results for 
thick-target bremsstrahlung model in Orbit 1; bottom panel shows results for Orbit 2.
Dashed lines indicate equivalent width measurements from Table~\ref{tbl:fits}.
Crosses indicate values for best-fit $\delta$, F$_{0}$ from spectral fitting
(Table~\ref{tbl:fits}), as well as maximum and minimum values using uncertainties
in $\delta$ and F$_{0}$.
\label{fig:kalpha}}
\end{center}
\end{figure}

\end{document}